%% file: Main.tex
\documentclass[manuscript,screen,acmsmall]{acmart}
\AtBeginDocument{%
  }

\setcopyright{acmlicensed}
\copyrightyear{2026}
\acmYear{2026}
\acmDOI{XXXXXXX.XXXXXXX}
\acmConference[CSCW 2026]{ACM Conference on Computer-Supported Cooperative Work and Social Computing}{October 10--14,
  2026}{Utah, USA}
\acmISBN{978-1-4503-XXXX-X/2018/06}

\begin{document}

\title{"When I see Jodie, I feel relaxed": Examining the Impact of a Virtual Supporter in Remote Psychotherapy \textit{(Preprint)}}

\author{Jiashuo Cao}
\affiliation{%
  \institution{The University of Auckland}
  \city{Auckland}
  \country{New Zealand}}
\email{jcao403@aucklanduni.ac.nz}

\author{Chen Li}
\affiliation{%
  \institution{The Hong Kong Polytechnic University}
  \city{Hong Kong}
  \country{Hong Kong}}
\email{richard-chen.li@polyu.edu.hk}

\author{Wujie Gao}
\affiliation{%
  \institution{The Hong Kong Polytechnic University}
  \city{Hong Kong}
  \country{Hong Kong}}
\email{ginawujie.gao@connect.polyu.hk}

\author{Simon Hoermann}
\affiliation{%
  \institution{University of Canterbury}
  \city{Christchurch}
  \country{New Zealand}}
\email{simon.hoermann@canterbury.ac.nz}

\author{Nilufar Baghaei}
\affiliation{%
  \institution{University of Queensland}
  \city{Brisbane}
  \country{Australia}}
\email{n.baghaei@uq.edu.au}

\author{Mark Billinghurst}
\affiliation{%
  \institution{The University of Auckland}
  \city{Auckland}
  \country{New Zealand}}
\email{mark.billinghurst@auckland.ac.nz}


\begin{abstract}

Virtual agents have shown promising potential in mental health applications, but current research has predominantly focused on contexts outside of traditional therapy sessions. This paper examines the impact of a virtual supporter in remote psychotherapy sessions conducted via Zoom. We used a two-phase research approach. First we conducted a formative study to understand the roles and functions of human supporters in psychotherapy contexts. Based on these findings, we developed a virtual supporter operating in two modes: Daily Mode (for mood journaling outside therapy) and Therapy Mode (as an additional participant in Zoom therapy sessions). Finally we ran a user study with 14 participants who engaged with the virtual supporter for a week and then joined a remote psychotherapy session together. Our findings revealed that the virtual supporter had positive effects on creating psychological safety, reducing anxiety, and enhancing emotional articulation without disrupting the therapeutic process. We then discussed both the benefits and potential disadvantages of virtual supporters in therapeutic contexts, including concerns about over-reliance and the need for appropriate boundaries. This research contributes to understanding how AI-driven virtual agents could contribute to human-led remote psychotherapy.
\end{abstract}




\maketitle

\section{Introduction}
\input{Sections/1-Introduction}

\section{Related Work}
\input{Sections/2-RelatedWork}

\section{Design Phase: From Therapist Insights to System Implementation}
\input{Sections/3-FormativeStudy}


\section{Evaluation Phase: User Study}
\input{Sections/6-Evaluation}

\section{Results}
\input{Sections/7-Results}

\section{Discussion}
\input{Sections/8-Discussion_Implication}

\section{Limitations and Future Work}
\input{Sections/9-Limitation_FutureWork}

\section{Conclusion}

\input{Sections/10-Conclusion}

\begin{acks}
To Robert, for the bagels and explaining CMYK and color spaces.
\end{acks}

\bibliographystyle{ACM-Reference-Format}
\bibliography{Reference}

\appendix
\section{Formative Study Interview} \label{appendix-interviewFormative}
As this is a semi-structured interview format, these questions serve as initial guidelines rather than a rigid script. The actual flow of conversation and participant responses will guide subsequent questions and their specific wording.
\subsection{General Unstructured Questions}
\begin{itemize}
    \item What does supporter mean to you?
    \item How do you understand the word "virtual agent"?
    \item Do you have any questions for me about this research?
    \item Is there anything you want to add?
\end{itemize}
\subsection{Experience with Human Supporters}
\begin{itemize}
    \item Could you describe your experience working with supporters in therapy sessions?
        \subitem How frequently do clients bring supporters?
        \subitem How do you feel when supporter attending therapy session?
    \item What are your observations about how supporters influence therapy sessions?
        \subitem Could you share a specific example of a positive influence?
        \subitem Have you encountered any challenges?
    \item How do you manage the dynamics between client, supporter, and you?
        \subitem Will you establish some restriction ahead?
\end{itemize}
\subsection{Virtual Supporters for Remote Therapy}
\begin{itemize}

    \item What potential benefits do you see in using virtual supporters in remote therapy?
        \subitem For which types of clients might this be most helpful?
        \subitem In what therapeutic contexts could this be valuable?
    \item What concerns do you have about incorporating virtual supporters?
        \subitem How might these concerns be addressed?
    \item How do you think virtual supporters might affect the therapeutic relationship?
        \subitem How might it influence the therapy dynamics?
        \subitem What role boundaries would be important?
\end{itemize}

\section{Therapy Session Procedure} \label{appendix-therapy}
\subsection{Introduction}
\begin{enumerate}
    \item Introduce the counselor and the purpose of the Zoom session.
    \item Confidentiality Statement: "The Zoom session will not be recorded, your privacy and comfort are important to me throughout this process."
    \item Session Structure and Expectations: "Since we only have 15-20 minutes, I may not be able to meet all your requests, but you have the right to interrupt the session and please feel free to talk anything you want."
    \item Ask if the client has any special needs or other concerns.
\end{enumerate}
\subsection{Engagement and Exploration}
\begin{enumerate}
    \item Initial Exploration: "Could you please tell me what bring you here? Why are you interested in our programme?"
    \item Problem Identification: "Could you please tell me what are yours presenting problems? Can you give me some examples? How do you feel about them?"
    \item Client Perspective: "What is your perception or understanding of the problems or concerns?"
\end{enumerate}
In this stage, the counselor will act as a mirror to reflect client's content. If time permits, the counselor will invite the client to talk more and discuss strengths and resources.

\subsection{Closure}
\begin{enumerate}
    \item Briefly summarize key points discussed. 
    \item Acknowledge the client's openness and participation.
    \item Transition Statement: "Thank you for sharing your experiences with me today. At this point, Jodie will guide you through a brief reflection about our session."
\end{enumerate}

\section{Evaluation Interview} \label{appendix-evaluation}
Similar to formative study interview, the question listed here mainly serve as guideline during the conversation.
\subsection{Questions for Participants}
\textbf{Overall Impressions of Jodie}
\begin{itemize}
    \item How would you describe your overall experience with Jodie?
    \item How would you describe your relationship with Jodie?
    \subitem Did this relationship change over the course of the study? If so, how?
    \item If you had to describe your relationship with Jodie using three words or phrases, what would they be?
\end{itemize}
\textbf{Daily Mood Journaling Experience}
\begin{itemize}
    \item What a typical mood journaling session with Jodie was like for you?
        \subitem What did you find most valuable/frustrate about these daily interactions?
    \item Did you notice any changes in how you interacted with Jodie during mood journaling over the course of the study?
    \item How comfortable did you feel sharing personal or emotional information with Jodie?
\end{itemize}
\textbf{Therapy Session Experience}
\begin{itemize}
    \item How did you feel about Jodie's presence during your therapy session?
    \subitem Did you notice Jodie during the therapy session? If so, at what moments?
    \item How will you describe the pre-therapy check-in/post-therapy reflection with Jodie?
    \item If you were to continue with therapy sessions in the future, would you want Jodie present? Why or why not?
\end{itemize}
\textbf{Suggestions for Improvement}
\begin{itemize}
    \item If you could change anything about Jodie, what would it be?
    \item What additional features or capabilities would make Jodie more helpful as a virtual supporter?
\end{itemize}

\subsection{Questions for Therapist}
\textbf{Overall Therapy Experience}
\begin{itemize}
    \item How would you describe the experience of conducting therapy sessions with Jodie present?
    \item Were there moments when you were particularly aware of Jodie's presence? If so, when and why?
    \item How did the technical implementation of Jodie as a Zoom participant work from your perspective?
\end{itemize}
\textbf{Comparison to Traditional Remote Therapy}
\begin{itemize}
    \item How did these therapy sessions compare to your typical remote therapy sessions without a virtual supporter?
    \item Did you notice any differences in client behavior in these sessions compared to typical first sessions without a virtual supporter?
    \item How did having Jodie present affect your own experience and approach as a therapist compared to standard remote sessions?
\end{itemize}

\section{Data Privacy and Usage Information Statement} \label{appendix-dataprivacy}
This appendix contains the section from our Participant Information Sheet that explained the data usage and privacy, which was provided and explained to participants during the study briefing phase. All participants received this information both verbally (through researcher explanation with slides) and in writing before beginning any interaction with Jodie.

\textbf{Data Collection and Storage}: During the study, we collect two types of data: 
(1) conversation logs from Jodie's interactions with participants during daily mood 
journaling and therapy sessions, stored through Soul Machines' platform and Dialogflow 
CX backend; and (2) questionnaire responses and interview recordings collected during the 
feedback phase. All data will be stored on secure, password-protected servers with 
encryption both in transit and at rest. Your data would not shared with Soul Machines, Google (Dialogflow CX), or any third-party commercial entities. All conversation logs were accessed through API calls and stored on our secure research servers, with no data retention by the commercial platforms beyond the technical requirements for real-time system operation. Access was restricted to members of the research team who had completed IRB-required training on human subjects research and data protection.

\textbf{Data Usage}: The data collected in this study will be used solely for research purposes. Jodie's dialogue system was built using pre-existing Soul Machines HumanOS capabilities and Dialogflow CX with fixed, rule-based dialogue flows that were established before participant recruitment began. No participant data will be used to modify, improve, or train Jodie's conversational capabilities during or after the study. Information you share with Jodie is for research purposes only and will NOT be used to provide clinical diagnosis, treatment recommendations, or mental health assessments outside the single therapy session included in this study. Researchers will analyze your mood journaling conversations, therapy session recordings, questionnaire responses, and interview transcripts to understand how virtual supporters affect therapy experiences. De-identified quotes from your conversations and interviews may be included in academic publications and presentations. Your name and any identifying details will be removed.

\textbf{Participant Rights on Data}: During the study, you can decline to answer any question, skip any questionnaire item, or refuse to discuss any topic without penalty. You can request to see any data we have collected about you at any time during or after the study by emailing the research team. If you believe any data we have recorded about you is inaccurate, you can request corrections within 30 days after completing the study by contacting the research team. You can withdraw from this study at any time without penalty or loss of benefits. If you withdraw, we will stop collecting new data from you immediately.
You can request that we delete all data collected from you up to that point. You can request deletion of your data at any time during the study or within 30 days after completing the study. After this 30-day window, your data will be fully anonymized (all identifiers permanently removed) and cannot be extracted from the research dataset.

\textbf{Data Anonymization}: To protect your privacy in research publications and presentations, we will replace your name with a participant code (P01, P02, etc.) in all transcripts and analysis documents Any identifying details mentioned during conversations or interviews-such as names, specific locations, workplace details, or other potentially 
identifying information-will be redacted from transcripts. The linking key that connects your participant ID to your real identity will be stored in a separate encrypted file, accessible only to the researcher. After the 30-day post-study window, the linking key will be permanently destroyed, making it impossible to connect the research data back to you.

\end{document}

%% file: Sections/1-Introduction.tex
This paper explores how a virtual supporter could be used to enhance remote psychotherapy sessions. Nearly one in eight people worldwide struggles with mental disorders, making mental health challenges a widespread global issue \cite{WorldHealthOrganization_2022}.  As demand for psychological services increases, technological advancements have enabled remote psychotherapy to become more popular. This shift encompasses both the transition from traditional face-to-face therapy to teletherapy via platforms like Zoom \cite{burgoyne2020lessons}, as well as the exploration of novel approaches to remote mental healthcare delivery \cite{easton2019virtual, provoost2017embodied}.

Within this evolving landscape, the use of virtual agents to provide mental health services represents an actively explored area \cite{he2023conversational}. Current research primarily focuses on employing virtual agents for emotion regulation activities, such as guided meditation \cite{hudlicka2013virtual} and mindfulness practices \cite{shamekhi_breathe_2015}, and for delivering specialized services, including virtual therapists or assessment interviews \cite{luerssen2018virtual}. While virtual agents have demonstrated promising capabilities, their use case scenarios have predominantly aimed at reducing or replacing the role of human therapists. A relatively underexplored domain is how virtual agents might enhance rather than replace traditional therapeutic settings-specifically, working alongside the therapist in human therapist-led psychotherapy sessions. 


Human therapist-led psychotherapy sessions typically involve only the client and therapist as participants. However, in some cases, a third person (e.g. trusted friend or paid carer) may join the therapy as a supporter for the client \cite{Meuleman2022Involving, scott2019supporting}. These supporters have been shown to provide crucial assistance to clients in therapeutic settings \cite{haydon2010want}. For instance, their presence creates notably positive experiences for clients with intellectual disabilities, facilitating more effective therapy participation, enhancing communication quality, and strengthening the therapeutic alliance between the client and therapist \cite{scott2019supporting}. The documented benefits of human supporters in therapy contexts point to a promising yet largely unexplored opportunity for developing virtual agents that could fulfill similar supportive functions.

Inspired by these benefits of human supporters in therapy sessions, we first conducted a formative study with nine therapists to gain deeper insights into the role of supporters in therapy sessions. Based on our findings, we designed and developed Jodie, a virtual supporter capable of accompanying users in remote psychotherapy sessions, and validated its effectiveness and impact through a user study with 14 participants. To summarise, our work makes the following contributions:
\begin{itemize}
    \item  Through interviews with practicing psychotherapists, we identified and characterized the primary roles of supporters in therapy contexts (comforter, information provider, and observer) along with their communication capabilities and boundary challenges.
    \item We developed a dual-mode virtual supporter system that operates both in daily contexts (through mobile-based mood journaling) and therapy contexts (as a Zoom participant), creating a continuous supportive presence that bridges between therapy sessions.
    \item We present one of the first empirical studies exploring the integration of virtual supporters in human-led remote psychotherapy settings, demonstrating their effectiveness in fulfilling supportive roles.
\end{itemize}

The main novelty of this work is that it is one of the first papers that presents a supportive virtual character designed from therapist feedback, and validated through a user study. This research contributes to understanding how AI-driven virtual agents could contribute to human-led remote psychotherapy, and so could lead to significant impacts in the future.

In the next section, we describe in more detail previous related work, and the research gap that we are addressing. Section 3 presents the design phase of our work, a formative study capturing feedback from therapists to inform the virtual agent design. We then describe the design and development of Jodie, the virtual supporter. Section 4 reports on a user study with the character, and Section 5 reports on the results of the user study. In Section 6, we discuss the results, while Section 7 highlights the limitations and directions for future work, with overall conclusions in Section 8. 


%% file: Sections/2-RelatedWork.tex
\subsection{The Use of Human Supporters in Psychotherapy}
In the context of therapy, a "Human Supporter" refers to a person who provides emotional and practical support to someone undergoing therapy, which could also be referred to as "Significant Other" \cite{Meuleman2022Involving}. The integration of human supporters in psychotherapy has emerged as a significant area of research and clinical practice \cite{keitner1990family, goodtherapyOtherPeople}, with substantial evidence pointing to its effectiveness in improving treatment outcomes across various mental health conditions \cite{ariss2020effect, Meuleman2022Involving, Jiménez-Murcia2017The}.

The presence of supporters in the treatment of post-traumatic stress disorder (PTSD) has been shown to reduce dropout rates and increase clients' motivation to begin therapy \cite{Meuleman2022Involving}. Research on gambling disorder treatment reveals similar patterns, with significantly lower dropout rates in groups where significant others attended therapy sessions alongside clients compared to those where clients participated alone \cite{Jiménez-Murcia2017The}. This benefit extends to group therapy settings as well; Dodding et al.\cite{{dodding2008all}} found that when participants brought supporters into group therapy, it led to improved outcomes in resilience, psychological health, and living environment. Looking at the supporter perspective, Scott et al.\cite{scott2019supporting} examined the experiences of those supporting individuals with intellectual disabilities during treatment and discovered that most supporters believed their involvement not only enhanced treatment effectiveness but also strengthened their relationships with the clients. Our research takes a distinct approach from previous studies by specifically examining the supporter's role from the therapist's perspective, providing a more nuanced understanding of the advantages and challenges that human supporters may introduce to therapeutic sessions.


\subsection{Virtual Agents in Mental Health}
Virtual agents have found widespread application across various psychotherapy contexts, from diagnosing disorders to delivering therapeutic services \cite{berube2021voice, vaidyam2019chatbots, cao2026digital}. The predominant research model features one-on-one interactions between virtual agents and participants, mirroring traditional therapeutic relationships \cite{Torous2018Empowering}. This approach has proven valuable in multiple clinical scenarios, as evidenced by Philip et al., who implemented a virtual agent as an interviewer conducting one-on-one dialogues with participants about depression symptoms \cite{philip2017virtual}. Similarly, Luerssen et al. demonstrated the efficacy of this model by utilizing a virtual agent as a therapist delivering Low-intensity Cognitive Behavioral Therapy (LiCBT)\cite{waller2013low} directly to participants through a mobile application, highlighting the potential for technology to extend therapeutic reach beyond traditional clinical settings \cite{luerssen2018virtual}.

A notable exception to this established paradigm appears in Lee et al.'s work \cite{lee2020designing}, where researchers designed a chatbot functioning as a mediator to facilitate deeper self-disclosure when participants interact with human therapists. However, even in this innovative approach, the chatbot's interactions remained exclusively participant-focused, maintaining the characteristic one-to-one interaction paradigm that dominates the field. This limitation reveals a significant gap in current research-the exploration of virtual agents in multi-party therapeutic contexts. Our research investigates the potential role of virtual supporters in scenarios where both participant and therapist are simultaneously present, potentially transforming the dynamics of therapeutic interactions and opening new possibilities for technology-enhanced mental healthcare delivery.



\subsection{Technology Support for Remote Psychotherapy}
Remote psychotherapy involves the use of digital technologies, such as video conferencing platforms like Zoom\footnote{https://www.zoom.com/} and Teams\footnote{https://teams.live.com/}, to conduct synchronous clinical therapy sessions between therapists and clients who are not physically co-located. Research has demonstrated that remote psychotherapy can achieve therapeutic outcomes comparable to traditional face-to-face therapy \cite{lin2022efficacy, giovanetti2022teletherapy}, making it an increasingly viable alternative for mental health service delivery.
However, remote psychotherapy faces distinct challenges. Some challenges mirror those found in traditional therapy, such as client anxiety and difficulty with self-disclosure due to unfamiliarity with the therapist \cite{ardito2011therapeutic, robledo2021therapy}. Others are unique to the technological medium, including the reduction of non-verbal cues that can weaken therapeutic alliance \cite{robledo2021therapy}.

Several HCI researchers have explored technological interventions to address these limitations. Lan et al. \cite{gao2025breaking} investigated the design space for sensing and sharing non-verbal cues in remote psychotherapy, recognizing the potential benefits and concerns the sensing technology could introduce to therapeutic practice. Hayoun et al. \cite{noh2024investigating} developed virtual masking technologies that provide clients with visual anonymity during remote therapy, demonstrating that increased privacy can lead to higher levels of self-disclosure. While our research takes a different approach by drawing inspiration from the documented benefits of human supporters in face-to-face therapy settings. Rather than attempting to replace human elements or compensate for technological limitations, we explore how virtual supporters can enhance the remote therapeutic environment by replicating the supportive functions that trusted companions provide in traditional therapy contexts.


%% file: Sections/3-FormativeStudy.tex
To inform the design of Jodie, our virtual supporter for remote psychotherapy, 
we conducted a two-stage process. First, we interviewed practicing psychotherapists 
to understand the roles, behaviors, and challenges of human supporters in therapy 
(Section 3.1-3.3). These insights directly shaped our design goals (Section 3.4), 
which we then operationalized through Jodie's system architecture, features, and 
implementation (Sections 3.5-3.7). This integrated approach ensures that our 
virtual supporter design is grounded in clinical expertise and therapeutic practice.

\subsection{Study Participants}
To be eligible for participation in the formative study, therapists were required to meet the following criteria: (1) hold a valid professional license to practice psychotherapy (e.g., Licensed Professional Counselor, Licensed Clinical Social Worker, Licensed Marriage and Family Therapist, or Licensed Psychologist); (2) have at least two years of post-licensure clinical experience; (3) have prior experience conducting therapy sessions with supporters present. After six weeks of recruitment through postings on a professional psychotherapy network and direct outreach, we recruited nine licensed psychotherapists (7 female, 2 male) aged 30 to 54 years. The participants represented diverse ethnic backgrounds: four identified as White, three as Asian, one as Hispanic or Latino/a/x, and one as Pacific Islander. All participants had prior experience with supporters joining therapy sessions. Notably, T8 and T9 had additional professional experience as caregivers for depressed client (four and three years, respectively). The therapists employed various therapeutic protocols in their practice, including Cognitive Behavioral Therapy (CBT)\cite{ellis1962reason,beck1970cognitive}, Acceptance and Commitment Therapy (ACT) \cite{hayes2005acceptance}, Dialectical Behavior Therapy (DBT) \cite{lynch2006mechanisms}, Solution-Focused Brief Therapy (SFBT) \cite{de2021more}, Emotion-Focused Therapy (EFT) \cite{greenberg2004emotion}, Interpersonal Psychotherapy (IPT) \cite{weissman2008comprehensive}, Art Therapy \cite{malchiodi2011handbook}, and Drama Therapy \cite{landy1994drama}.

\begin{table}[bp]
\begin{tabular}{|l|l|l|l|l|l|}
\hline
Participant & Age & Gender & Year of experience & Ethnicity              & Protocol           \\ \hline
T1          & 30  & Female & 3                  & Asian                  & CBT, SFBT          \\ \hline
T2          & 52  & Male   & 15                 & White                 & CBT, ACT           \\ \hline
T3          & 54  & Female & 9                  & White                 & CBT, EFT           \\ \hline
T4          & 44  & Female & 6                  & White                 & IPT                \\ \hline
T5          & 34  & Female & 2                  & Pacific Islander               & ACT                \\ \hline
T6          & 30  & Male   & 5                  & Asian                  & CBT, Drama Therapy \\ \hline
T7          & 46  & Female & 4                  & Hispanic or Latino/a/x & ACT                \\ \hline
T8          & 42  & Female & 11                 & Asian                  & Art Therapy        \\ \hline
T9          & 39  & Female & 7                  & White                 & CBT, DBT           \\ \hline
\end{tabular}
\caption{Overview of the psychotherapists taking part in the study}
    \label{tab:my_label}
\end{table}

\subsection{Study Procedure}
Prior to the interviews, each participant completed a demographic questionnaire and engaged in a brief intake conversation to establish rapport. Recognizing that participants had varying levels of familiarity with virtual agents, we decided to include a demonstration of a virtual agent during the interviews to facilitate more concrete discussion about the potential role of virtual supporters. We selected embodied agents from Soul Machines\footnote{https://www.soulmachines.com/} with realistic appearances and behaviors, as support persons traditionally represent roles fulfilled by humans in therapeutic settings.
The semi-structured interviews were conducted via Zoom and were divided into three parts (see Appendix \ref{appendix-interviewFormative}): 1. Current Support Practices: Questions about therapists' observations and thoughts regarding human supporters during their therapeutic practice, including typical roles, behaviors, and challenges.
2. Virtual Agent Demonstration: Presentation of a virtual agent's capabilities, including dialogue examples and non-verbal behaviors, to provide participants with a concrete reference for discussion.
3. Expectations and Concerns: Exploration of therapists' expectations, requirements, and potential concerns regarding implementing a virtual supporter in therapy sessions.

Interview durations ranged from 37 to 61 minutes (mean=47.2, SD=7.93). All sessions were recorded with participant consent. Following data collection, we performed thematic analysis\cite{braun2006using} on the interview transcripts. First, we transcribed all interviews and systematically analyzed the transcripts to generate initial codes, identifying meaningful patterns in the data. After coding completion, we organized these codes into five broad categories: 1) benefits of human supporters, 2) challenges of human supporters, 3) potential of virtual supporters, 4) concerns about virtual supporters, and 5) attitudes towards human versus virtual supporters. Based on these categories and the underlying codes, we developed themes through an iterative process of reviewing and refining. This process ensured that the final themes accurately captured the key patterns and insights from the interviews, providing a comprehensive understanding of therapists' perspectives on both human and virtual supporters in therapeutic contexts. The final thematic structure was established after reaching consensus among the research team.

\subsection{Study Findings}
Our analysis revealed four main themes related to the design requirements for virtual supporters in psychotherapy: (1) Therapeutic Roles and Functions, (2) Communication Capabilities, (3) Relationship Dynamics and Boundaries and (4) Considerations on Virtual Supporter. Below, we elaborate on each theme with supporting evidence from our interviews.

\subsubsection{Therapeutic Roles and Functions}
Based on therapists' descriptions of supporter behaviors, we categorized supporters' specific roles and functions into three types: comforter, information provider, and observer.

\textbf{Supporter as Comforter: }
Participants reported that clients demonstrated increased comfort and confidence with supporters present during therapy sessions. T4 noted that a supporter's presence was instrumental in \textit{"creating a safer environment... for the client."} This enhanced comfort stemmed from the social relationships supporters share with clients, characterized by familiarity and mutual trust \cite{cohen2004social}. The comfort-enhancing effect was particularly significant during the first sessions, when both the therapist and therapy environment were unfamiliar. During these initial encounters, supporters helped mitigate clients' anxiety and feelings of insecurity arising from the novelty of the setting.
Beyond the comforting effect of their mere presence, supporters often engaged in specific behaviors to enhance client comfort during therapy sessions. These behaviors included active listening (maintaining eye contact, nodding, etc.), which not only made clients feel heard but also provided affirmation during their narratives. As T7 explained, \textit{"They (clients) sometimes look toward the supporter when they're uncertain about something, and at these moments, the supporter's eye contact and nodding can give clients great confidence to continue their narrative."} Such validation from supporters effectively addressed clients' self-doubt, facilitated session progress, and strengthened clients' confidence. Additionally, when clients experienced significant emotional fluctuations, supporters would often provide physical comfort through touch, such as gently stroking their shoulders and backs, or holding their hands to offer support.

\textbf{Supporter as Information Provider: }
Supporters sometimes joined conversations between clients and therapists at the client's invitation. This typically occurred in two situations. First, when clients were uncertain about the accuracy of their narratives, they would seek confirmation or correction from their supporters. T2 described a common interaction pattern: \textit{"...they (clients) might turn around to the support person and ask them about it. 'You remember this is what happened, right?' or 'is that correct' or 'do you think that am I missing something here?'"}
The second situation arose when supporters themselves had experienced the events being discussed. Clients would actively invite supporters to join in the narrative. One observed behavior, as noted by T3, was that \textit{"sometimes they'll like finish the sentence for them (clients) sometimes."}

\textbf{Supporter as Observer:}
Supporters in therapy sessions did more than simply provide help to clients; they continuously observed them, an observation that extended into daily interactions. As T8 stated, \textit{"As a professional carer, one of my important jobs is to observe my client. This not only helps me identify changes in their needs promptly but also allows me to better understand their emotional states, preferences, and lifestyle habits, thereby building a stronger trust relationship."}
Beyond helping supporters establish deeper mutual trust with clients, these accumulated observations formed an alternative perspective that could assist clients when needed. One example was post-session reflection discussions, as described by T9: \textit{"(As a professional carer) I sometimes chat with my client after they complete a session, discussing their feelings, and they also ask for my thoughts."}

\subsubsection{Communication Capabilities}
Our analysis found that effective supporters demonstrate sophisticated communication capabilities that facilitate therapeutic progress. These capabilities fall into three main categories: active listening behaviors, appropriate verbal interventions, and nonverbal communication skills.

\textbf{Active Listening Behaviors:}
Therapists consistently emphasized the importance of active listening behaviors exhibited by supporters. These behaviors included maintaining appropriate eye contact, nodding at suitable moments, and displaying facial expressions that conveyed understanding and empathy. T6 explained: \textit{"The helpful supporters know when to just be present and listen. Their facial expressions and body language tell the client, 'I'm here, I'm following what you're saying, and I understand,' without needing to interrupt the session."}
This type of attentive presence created what T8 called "a witnessing space" that validated clients' experiences and encouraged deeper emotional disclosure. Several therapists noted that clients would often look to their supporters for nonverbal feedback during challenging narratives, seeking confirmation through subtle glances that were met with affirming nods or expressions.

\textbf{Appropriate Verbal Interventions:}
While supporters' nonverbal presence was crucial, therapists also identified patterns of effective verbal communication. Successful supporters demonstrated an intuitive understanding of when to speak and when to remain silent. T2 observed: \textit{"Good supporters don't actually talk for a long time... they only join the conversation when necessary, and most of the time they're actually sitting quietly there."}
When supporters did speak, therapists noted that the most helpful contributions were brief, supportive statements that provided clarification or gentle encouragement rather than dominating the conversation. The common phrases that supporter uses are like "yeah you're right" or "I remember you mentioned..." that add context without taking over the client's narrative. Several therapists mentioned that they think a good supporter should avoid offering their own solutions or interpretations during therapy sessions that might confuse clients' thoughts and undermine the therapist's approach.

\textbf{Nonverbal Communication Skills:}
Beyond facial expressions and listening behaviors, therapists highlighted the importance of body positioning, proximity, and touch in supporter communication. T7 noted that supporters often positioned themselves slightly behind or beside clients, creating a physical arrangement that communicated support without dominance, this finding got repeated by T8, saying \textit{"They (supporter) usually stay in the background. But it also depends on where the seat [is] in the session, they might sit between the client and the therapist on the side."} This positioning allowed clients to maintain primary eye contact with therapists while still sensing the supporter's presence.
For clients experiencing emotional distress, appropriate touch was identified as a powerful communication tool unavailable to therapists due to professional boundaries. T2 described: \textit{"When clients become overwhelmed, a supporter can offer a hand on the shoulder or arm, and that simple contact can ground someone enough to continue the therapeutic work." }

\subsubsection{Relationship Dynamics and Boundary}
There are two significant challenges in the relationship dynamics between clients, supporters, and therapists: over-reliance and excessive support, and role boundary navigation. These findings highlight the complex nature of incorporating supporters into therapeutic contexts.

\textbf{Over-reliance and Excessive Support:}
Therapists expressed concerns about clients potentially developing over-reliance on their supporters during therapy sessions. Several participants observed that while supporter presence initially facilitated comfort and confidence, some clients began to demonstrate excessive dependence on supporters for emotional regulation or decision-making. As therapy progressed, this dependence could potentially hinder the development of clients' autonomous coping skills.
T5 noted: \textit{"I probably [do] not want to see the client over-rely on the support person ... and I make it very clear that I'm not there to do couples counseling. We're all here for the client."}. 
Therapists also identified cases where supporters provided excessive support that inadvertently undermined therapeutic goals, including cases of oversharing personal information or making comments that conflicted with clients' expressed desires or lived experiences. As one therapist noted, \textit{"you always get the people that talk too much, and it's not about them. It's about the person who's having the therapy."} (T3). Both concerns, over-reliance and excessive support, fundamentally represent a diminishment of clients' agency, which is crucial for effective therapeutic outcomes and advancement \cite{williams2007principles}. As the primary focus of therapy, clients must engage actively in their treatment process; when supporters unintentionally diminish this agency, the therapeutic process becomes increasingly passive, potentially hindering the client's development and long-term acquisition of self-efficacy skills.

\textbf{Role Boundary Navigation:}
To prevent the challenges presented above, several therapists mentioned using explicit boundary-setting discussions at the beginning of treatment to establish expectations. T4 stated: \textit{"I like to have a clear conversation about roles with both the client and supporter present. We talk about when input would be helpful and when it might interfere with the therapeutic process."} These conversations included discussing confidentiality boundaries, particularly when supporters might occasionally be absent from sessions.
Another approach therapists mentioned involves actively intervening during sessions when they perceive problematic interaction dynamics between supporters and clients, particularly when the conversation's focus shifts away from the client. Therapists deliberately step into a mediator role to redirect and guide the dialogue. As T3 explained, \textit{"When I sense something isn't quite right, I turn into a mediator kind of thing, and remind everyone about what we've been working toward in our previous sessions."} This mediation strategy allows therapists to realign conversations with therapeutic objectives while preserving the collaborative environment, ensuring that client needs remain central to the session.

\subsubsection{Considerations on Virtual Supporter}
Therapists held nuanced perspectives regarding the potential of virtual supporters in therapeutic contexts. Their assessments were informed by their experiences with human supporters and reflected both enthusiasm about potential benefits and caution regarding implementation challenges. These considerations clustered around three main areas: perceived advantages of virtual supporters, concerns about implementation, and the presence of a virtual supporter.

\textbf{Perceived Advantages of a Virtual Supporter:}
Therapists identified several unique advantages that virtual supporters might offer compared to human supporters. First, they noted that virtual supporters could address some boundary issues identified when working with human supporters. By converting the boundary-setting phase into programmed behavioral and linguistic boundaries, virtual supporter could maintain consistent adherence to therapeutic protocols and avoid potentially harmful or disruptive behaviors, effectively replicating human supporter preparation but with greater reliability and less time spent.
Several therapists also highlighted the potential for greater accessibility and convenience. This was seen as particularly valuable for clients with limited social networks, such as international students and first-generation immigrants. Virtual supporters would be available regardless of time constraints, geographical limitations, or scheduling conflicts that often complicate human support arrangements.
Participants also anticipated that virtual supporters might eliminate some of the emotional complexities that arise with human supporters. As T2 mentioned: \textit{"With human supporters, there's always the relationship history and dynamics to consider. ... It might be easier to have someone (virtual supporter) that's only there for YOU and never judge you."} Therapists suggested this emotional neutrality and privacy might create a safer space for certain clients to express themselves without fear of judgment or causing distress to their supporter.

\textbf{Concerns About Implementation:}
Despite recognizing potential benefits, therapists expressed several concerns about implementing virtual supporters in therapy. A primary concern is related to the development of the relationship between clients and virtual supporters. Therapists emphasized that all supportive feelings are based on authentic, trustworthy relationships, making the pre-therapy relationship-building stage equally important for virtual supporters. T8 drew on previous professional career experience to suggest: \textit{"Perhaps one approach to establishing (a) connection you could try is completing mood journaling with the client before therapy begins."} This recommendation highlighted the importance of creating meaningful engagement prior to therapeutic work to ensure that supportive behaviors would be perceived as genuine.
Privacy was identified as another critical concern. Therapists emphasized that for virtual supporters (as digital entities), privacy would be best achieved through transparency - ensuring clients clearly understand what information is being collected and how it's being used. This transparency would enable clients to maintain a sense of control over their personal information and better establish trust with the virtual supporter. As T1 noted: \textit{"Do my clients know what is happening with the virtual supporter? ... It's all about keep[ing] the session private and safe."} Participants stressed that clear communication about data practices was fundamental to creating the safe environment necessary for virtual supporters to be effective.

\textbf{The Virtual Supporter Interface:}
Therapists discussed how a virtual supporter should be presented within the therapeutic environment. There was a strong consensus that Zoom, as the current mainstream platform for remote psychotherapy, would be the appropriate foundation for displaying the virtual supporter. Participants reasoned that since the virtual supporter itself represents a novel concept, pairing it with an unfamiliar platform would create excessive learning and cognitive burden for therapists, potentially hampering subsequent testing and evaluation.
During the second part of the interviews, when the virtual supporter was demonstrated, many therapists requested that researchers position the virtual supporter's upper body (shoulders and above) within the screen share. They later remarked that this presentation \textit{"almost felt like I was talking with a person (in Zoom)}" (T5). Consequently, when discussing display methods for the virtual supporter, most therapists advocated for simulating human-like presentation - specifically, having the virtual supporter appear as an independent video stream within the Zoom interface, similar to how human participants appear during video conferences.

\subsection{System Design Goals}
Based on our formative study findings, we established five primary design goals for Jodie, our virtual supporter for remote psychotherapy. These goals directly address the key insights from therapist interviews and provide a framework for Jodie's design.

\subsubsection{DG1: Building Towards Comforter and Observer Roles} \label{DG1} 
Our first design goal was to optimize Jodie for the two most valued supporter functions: providing comfort and maintaining observational continuity. Therapists consistently highlighted these roles as offering the greatest therapeutic benefit while presenting fewer boundary complications than information provision.

For the comforter role, we aimed to create a virtual presence that could provide emotional reassurance during therapy sessions through appropriate nonverbal behaviors. These subtle behaviors would be designed to enhance client security without interrupting the therapeutic dialogue.
For the observer role, our goal is to enable users to engage in everyday conversations and share their feelings across sessions, creating the continuity of care that professional supporters described as valuable. This longitudinal observation would allow Jodie to provide relevant support based on the client's evolving needs and the progression of therapy.


\subsubsection{DG2: Relationship Building with Clear Boundaries} \label{DG2}
Our second design goal was to facilitate authentic relationship development between clients and Jodie while maintaining appropriate boundaries that prevent over-reliance and preserve therapeutic agency. 
We aimed to create a support experience that balances two critical needs: building genuine connection through regular engagement, and preventing the over-reliance and diminishment of client agency that therapists identified as problematic with human supporters. Unlike human supporters who might inadvertently provide excessive support, we embedded intentional limitations and clear role boundaries into Jodie's design from the outset.

\subsubsection{DG3: Simulating Natural Human Interaction} \label{DG3}
Since our study focused on learning from human supporters, our central design goal was to create natural, human-like interaction experiences that would feel authentic and accessible to users across different contexts while reducing the cognitive burden of technology. We aimed to enable natural conversational engagement, allowing users to communicate as they would with a human supporter. We also prioritized providing seamless accessibility across different devices and contexts, ensuring users could engage with Jodie in ways that felt familiar and contextually appropriate to their environment.

\subsubsection{DG4: Integrate into Existing Therapeutic Workflows} \label{DG4}
Our fourth design goal was to minimize disruption to established remote therapy practices by embedding Jodie within familiar platforms that therapists and clients already use. Our goal was to reduce the learning curve and technical barriers associated with adopting new technology, allowing therapeutic professionals to focus on care delivery rather than system management.
By leveraging existing video conferencing infrastructure, we sought to ensure that Jodie would enhance the therapeutic process without introducing additional complexity or requiring significant workflow adjustments.

\subsubsection{DG5: Transparency and Privacy Protection} \label{DG5}
Our final design goal was to establish clear transparency and strong privacy protections, recognizing these as foundational to therapeutic trust. Rather than attempting to make data collection invisible, we aimed to make these processes explicitly visible and understandable to clients.
Recognizing that transparency is both an ethical imperative and a functional requirement for effective therapeutic relationships, we intended to create a system where privacy would be achieved through informed participation and client agency, enabling users to engage confidently with the technology.

\subsection{System Design Concept: Dual-Mode Embodied Conversational Agent}
Building on the design goals derived from our formative study, we developed Jodie 
as a dual-mode virtual supporter system. \textbf{Daily Mode} focuses on relationship building (DG2) and fulfills the observer role (DG1). In this mode, Jodie primarily
engages with users through daily conversational mood journaling, helping them record and reflect on their emotional states, creating continuity of care across therapy sessions. To simulate the intimacy of FaceTime conversations with friends (DG3), Daily Mode presents Jodie with close-up visibility, showing the face and upper shoulders prominently on screen (See Figure \ref{fig:jodie_use}a).

\textbf{Therapy Mode} is optimized for providing emotional support during actual therapy 
sessions, embodying the comforter role (DG1). In this mode, Jodie employs specific comforting behaviors, such as attentive listening, affirmative nodding, and appropriate facial expressions, to provide emotional reassurance without verbal interruption during therapy sessions. Jodie also engages with users before and after therapy sessions, helping them transition into and out of the therapeutic space through brief check-ins that facilitate emotional preparation and post-session reflection. To integrate seamlessly with Zoom, the predominant platform for remote therapy (DG4), we designed Jodie to appear as a participant within the video call, with visual framing that matches how human participants typically appear, shown from approximately chest-up in the standard gallery view (See Figure \ref{fig:jodie_use}b).

To maintain clear boundaries and prevent over-reliance (DG2), Jodie utilizes rule-based conversation dialogue across both modes. This architectural decision ensures that all conversational content and interaction flows are carefully controlled and therapeutically appropriate. The rule-based system prevents Jodie from overstepping into advice-giving or decision-making roles that could compromise client agency or therapeutic boundaries, addressing concerns therapists expressed about human supporters who inadvertently provided excessive support.

Addressing DG5, we established strict data handling procedures to protect participant privacy. All dialogue data was stored exclusively on our secure research servers, with no data retention by any commercial online platforms. We provided participants with full access to and control over their data. Participants could review their recorded data at the end of the study and request any modifications, corrections, or deletions. Additionally, we implemented a mandatory briefing process that occurred before users began interacting with Jodie. This briefing clearly explained all data collection and usage procedures, ensuring users understood exactly what information Jodie would access, how it would be used, what controls they had over their data, and confirmation that all participant data would be used solely for this research study. For full details of our data management protocol, please refer to Appendix \ref{appendix-dataprivacy}. 

To summarize, Table \ref{tab:design_mapping} shows how each design goal was translated into specific system features.

\begin{table}[h] 
\centering
\caption{Translation of design goals into system features}
\label{tab:design_mapping}
\begin{tabular}{p{3cm}|p{4.5cm}|p{6cm}}
\hline
\textbf{Design Goal} & \textbf{System Feature(s)} & \textbf{Implementation Details} \\
\hline
DG1: Comforter \& Observer Roles & 
\textbullet\ Pre-therapy check-in\newline
\textbullet\ Proper listening behaviors\newline
\textbullet\ Post-therapy reflection\newline
\textbullet\ Daily mood journaling & 
Emotional reassurance through nonverbal behaviors during therapy; longitudinal emotional tracking through daily conversations \\
\hline
DG2: Relationship Building with Clear Boundaries & 
\textbullet\ Daily mood journaling\newline
\textbullet\ Redirecting Off-Topic Dialogue & 
Structured daily engagement with rule-based boundaries to prevent over-reliance \\
\hline
DG3: Natural Interaction & 
\textbullet\ Near-human appearance\newline
\textbullet\ Natural language processing\newline
\textbullet\ Emotionally congruent expressions & 
Soul Machines HumanOS platform; conversational interface requiring no technical commands \\
\hline
DG4: Integrate into Existing Workflows & 
\textbullet\ Zoom integration for therapy\newline
\textbullet\ Mobile app for daily use & 
Jodie appears as participant in zoom calls; familiar mobile interface \\
\hline
DG5: Transparency \& Privacy & 
\textbullet\ Mandatory briefing process\newline
\textbullet\ Explicit data controls\newline
\textbullet\ Clear capability boundaries & 
Pre-use orientation; user control over information; honest limitation disclosure \\
\hline
\end{tabular}
\end{table}

\subsection{System Features}
In this section, we detail the specific features of Jodie implemented in Daily Mode and Therapy Mode. 

\subsubsection{General Features}
Jodie was developed based on Soul Machines HumanOS, which provided fundamental capabilities that aligned well with our DG3. This foundation equipped Jodie with several inherent features, including near-human appearance, natural language interaction capabilities, and emotionally congruent facial expressions and body language. These built-in capabilities provided an excellent starting point for creating a virtual supporter that could simulate natural human interaction while fulfilling the comforter and observer roles identified in our formative study.

\begin{figure}
    \centering
    \includegraphics[width=0.8\linewidth]{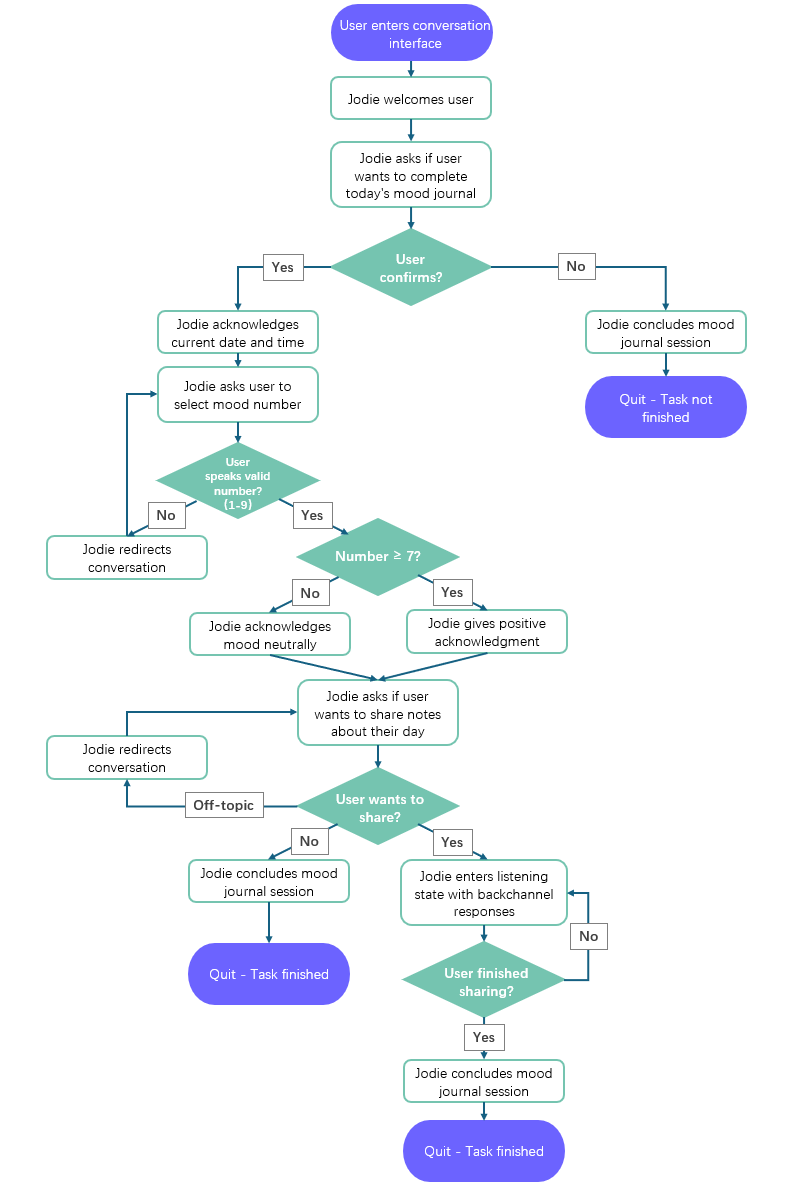}
    \caption{Daily Mode Workflow}
    \Description[Daily Mode Workflow]{A flowchart illustrating the interaction flow of the virtual supporter in a daily use context, comprising four stages: prompting the user to begin their mood journal for the day, asking the user to select a number from 1 to 9 to indicate their current mood, inviting the user to share any events or experiences they would like to talk about, and concluding the session.}
    \label{fig:daily_mode}
\end{figure}

\subsubsection{Daily Mode Features} \label{dailyModeFeature}

\textbf{Mood Journaling with User}: Inspired by the commonly used DBSA wellness tracker \footnote{https://www.dbsalliance.org/wp-content/uploads/2020/07/DBSA-WellnessTracker-Mood.pdf}, we developed a conversational approach to mood journaling (see Figure \ref{fig:daily_mode}). When users enter the conversation interface, Jodie first welcomes them and asks if they would like to complete today's mood journal. After the user confirms, Jodie acknowledges the current date and time (e.g., "Cool! Today is Monday and now is 9PM."). This verbal acknowledgment serves to record the temporal context of the mood rating, as the time of day when mood journaling occurs is an important factor in the DBSA wellness tracker framework. Jodie then asks the user to select a number from 1-9 (ranging from extremely depressed to extremely elevated mood) to represent their current emotional state. Users must vocalize their selection (e.g., "Right now is around 6") to advance to the next stage. If the user says a number greater than or equal to seven, Jodie responds with positive emotional acknowledgement (e.g., "Sounds like a good day!"). Jodie then asks if the user would like to share significant notes about their day (e.g., "Do you feel like talking about what went on today?"). When the user begins sharing, Jodie enters a listening state, performing listening behaviors and occasionally inserting backchannel responses (e.g., "right," "I see," "un huh") during pauses in the user's narrative. This continues until the user indicates they have finished sharing, at which point Jodie concludes the mood journal session and says goodbye.

\textbf{Redirecting Off-Topic Dialogue}: Due to the unpredictability of daily conversations, maintaining appropriate role boundaries through daily interaction required Jodie to actively manage conversation scope. In mood journaling, the conversation progresses only when Jodie receives user intentions that align with the expected dialogue flow. Jodie identifies user input as off-topic when it falls into the following categories: (1) responses that fail to provide a valid mood number (1-9) when prompted for mood rating, including non-numerical responses or numbers outside the specified range; or (2) responses to the sharing invitation that cannot be clearly classified as affirmative or negative (e.g., ambiguous statements or unrelated questions). When user dialogue is classified as off-topic, Jodie employs standardized redirect responses to guide the conversation back to the current mood journaling step (e.g., "Sorry, I don't understand that. Could we go back to what we were talking about before?"). There is no restriction on the number of retries allowed in Jodie's conversational system.

\begin{figure}
    \centering
    \includegraphics[width=\linewidth]{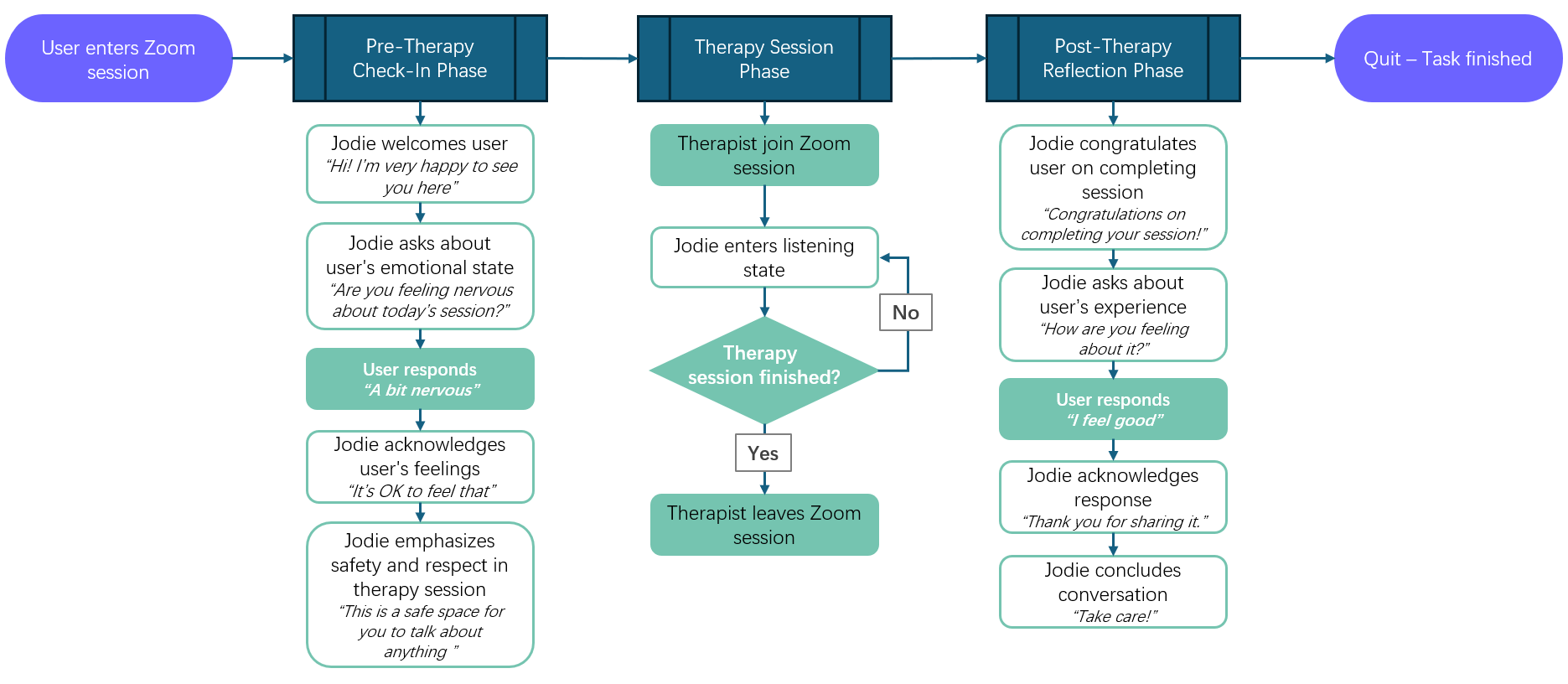}
    \caption{Therapy Mode workflow}
    \Description[Therapy Mode workflow]{A flowchart illustrating the interaction workflow of the virtual supporter within a remote therapy context, encompassing three phases: providing emotional comfort to the user prior to the therapy session, maintaining active listening behaviours throughout the therapy session, and facilitating reflective dialogue with the user following the completion of the session.}
    \label{fig:therapy_mode}
\end{figure}

\subsubsection{Therapy Mode Features} \label{therapyModeFeature}
\textbf{Pre-Therapy Check-In}: Before the therapist enters the Zoom session, Jodie engages in a brief conversation with the user (see Figure \ref{fig:therapy_mode}). Jodie expresses happiness about joining the therapy session and inquires about the user's current emotional state (e.g., "Hi! I'm very happy to see you here! Before the therapist arrives, I just wanted to check in with you. Are you feeling nervous about today's session?"). After the user responds, Jodie acknowledges their feelings and gently emphasizes the importance of safety and respect in therapy (e.g., "I just want to let you know that this is a safe space for you to talk about anything on your mind. There's no pressure to dive into anything you're not ready for.").

\textbf{Proper Listening Behaviour}: Once the therapist joins the Zoom session and therapy begins, Jodie enters a listening state. During this time, Jodie displays facial expressions corresponding to positive and negative emotions expressed in the conversation and periodically nods to signal attentiveness. Importantly, Jodie does not make any verbal responses until the therapist leaves the Zoom session, maintaining appropriate boundaries during the therapeutic dialogue.

\textbf{Post-Therapy Reflection}: After the therapy session concludes, Jodie congratulates the user on completing their therapy session and inquires about their experience, including their feelings about the session, impressions of the therapist, and any other thoughts they wish to share (e.g., "I see. Is there anything else you want to share about today's session?"). Finally, Jodie offers encouragement and concludes the conversation.

\subsection{System Implementation Details}
\begin{figure}
    \centering
    \includegraphics[width=\linewidth]{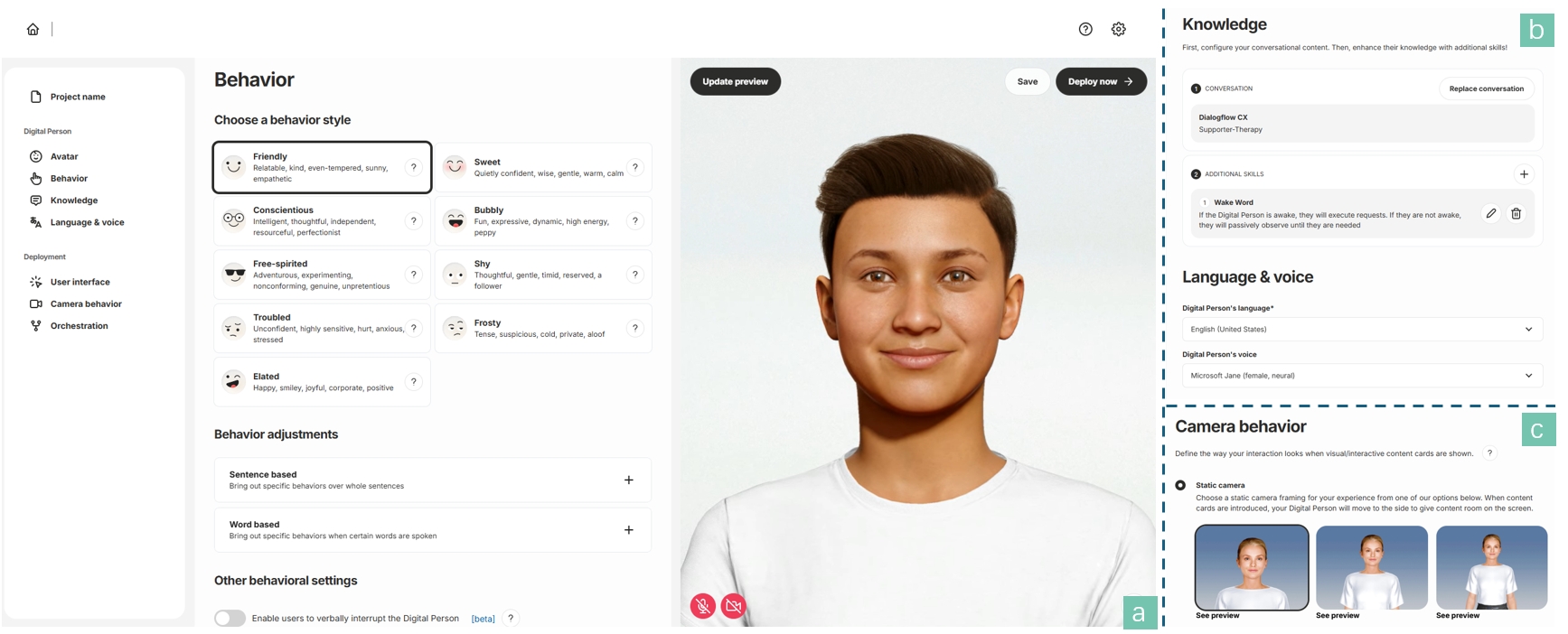}
    \caption{a: Overview of Soul Machines Studio. b: Setting for Knowledge (dialogue management) and Language \& voices. c: Setting for Camera behavior}
    \Description[Screenshots of Soul Machines Studio.]{A collection of Soul Machines Studio's screenshots, include a: Overview of Soul Machines Studio. b: Setting for Knowledge (dialogue management) and Language \& voices. c: Setting for Camera behavior}
    \label{fig:soul_machine_studio}
\end{figure}

Jodie was primarily developed using Soul Machines Studio (see Figure \ref{fig:soul_machine_studio}), a platform that enables the creation of virtual agents based on Soul Machines Human OS \footnote{https://soulmachines-support.atlassian.net/wiki/spaces/SSAS/pages/381353985/Human+OS}. The platform provides automated control of the agent's facial expressions and nonverbal behaviours based on speech content through Experiential AI\footnote{https://www.soulmachines.com/experiential-ai}, allowing us to focus on conversational design. After discussions with therapists regarding appropriate appearance, we selected the preset "Alba" appearance from the available options. This choice was deliberate as Alba's features do not exhibit pronounced racial characteristics, making her more universally acceptable across diverse populations. To align with the supporter role, we configured her behavioral style as "Friendly," which provided a baseline of warm, approachable, and non-judgmental interaction patterns.

For speech recognition, we utilized Deepgram's automatic speech recognition (ASR) service, which is natively integrated into Soul Machines Human OS. According to Kuhn et al.\cite{kuhn2024measuring}, the Deepgram ASR achieves an average word error rate (WER) of 8.3\% under standard operating conditions \cite{kuhn2024measuring}. The emotion recognition is also fulfilled by Deepgram API. For dialogue management, we implemented Dialogflow CX\footnote{https://dialogflow.cloud.google.com/cx} as the backend control system. This platform was selected for its robust intent recognition capabilities, flexible conversation flow management, and seamless integration with the Soul Machines ecosystem. Using Dialogflow CX allowed us to create structured conversation paths while maintaining natural-feeling interactions within defined therapeutic boundaries. For example, when Jodie asks the user whether they would like to share what happened today, the user is not required to respond with a solid 'Yes' or 'No'; instead, the system infers the user's communicative intent to advance the conversation. To reduce potential risks in the conversation, all dialogue content is proofed by the therapist. We also activated HumanOS's Wake Word functionality to manage Jodie's state transitions in Therapy Mode. This feature enables the agent to enter a "sleep mode" in which active interaction is stopped and only listening behaviors are displayed. The agent remains in this state until a predefined wake word—a specific phrase or sentence spoken by the user—triggers reactivation. In our implementation, Jodie transitioned to sleep mode upon completing the pre-therapy check-in conversation. To initiate the post-therapy reflection phase, the therapist was instructed to say "Thank you for coming today" before leaving the Zoom session. This phrase served as the wake word that prompted Jodie to resume active dialogue for the post-therapy reflection.

Regarding camera behavior settings, we configured Jodie with a static camera position using a head-and-shoulders framing. This visual presentation ensures consistency across different platforms while matching the typical appearance of human participants in video calls and Zoom sessions. The head and shoulders framing is particularly familiar to users accustomed to video conferencing, helping Jodie integrate naturally into the remote therapy environment. Based on our target audience for this study, we set Jodie's language to English.

Based on Soul Machines documentation \footnote{https://support.soulmachines.com/support/solutions/articles/101000534182-latency-and-speech-recognition}, the system latency is highly dependent on the user's network connection. With a stable connection, there is approximately 0.5 seconds of latency between when the user starts talking and when the agent performs listening behaviors (such as nodding or displaying attentive expressions). This latency represents the general response time for the system to process audio input and trigger appropriate nonverbal behaviors.

After completing development, we deployed Jodie as a web-based application using Soul Machines Studio's publishing functionality. To enable Jodie's participation in Zoom sessions as if she were a human participant, we implemented a technical workflow using OBS Studio\footnote{https://obsproject.com/} to capture Jodie's web interface in real-time. We then configured Zoom to use OBS Virtual Camera as its video source, effectively displaying Jodie within the Zoom environment alongside human participants. For audio functionality, we used VB-Audio software\footnote{https://vb-audio.com/} to create virtual audio input/output ports, setting Zoom's microphone and speaker to Virtual Cable Input/Output, respectively. This configuration created a complete audio loop, allowing Jodie to receive audio from the Zoom session and respond through Zoom's audio channel.

%% file: Sections/6-Evaluation.tex
\begin{figure}
    \centering
    \includegraphics[width=0.8\linewidth]{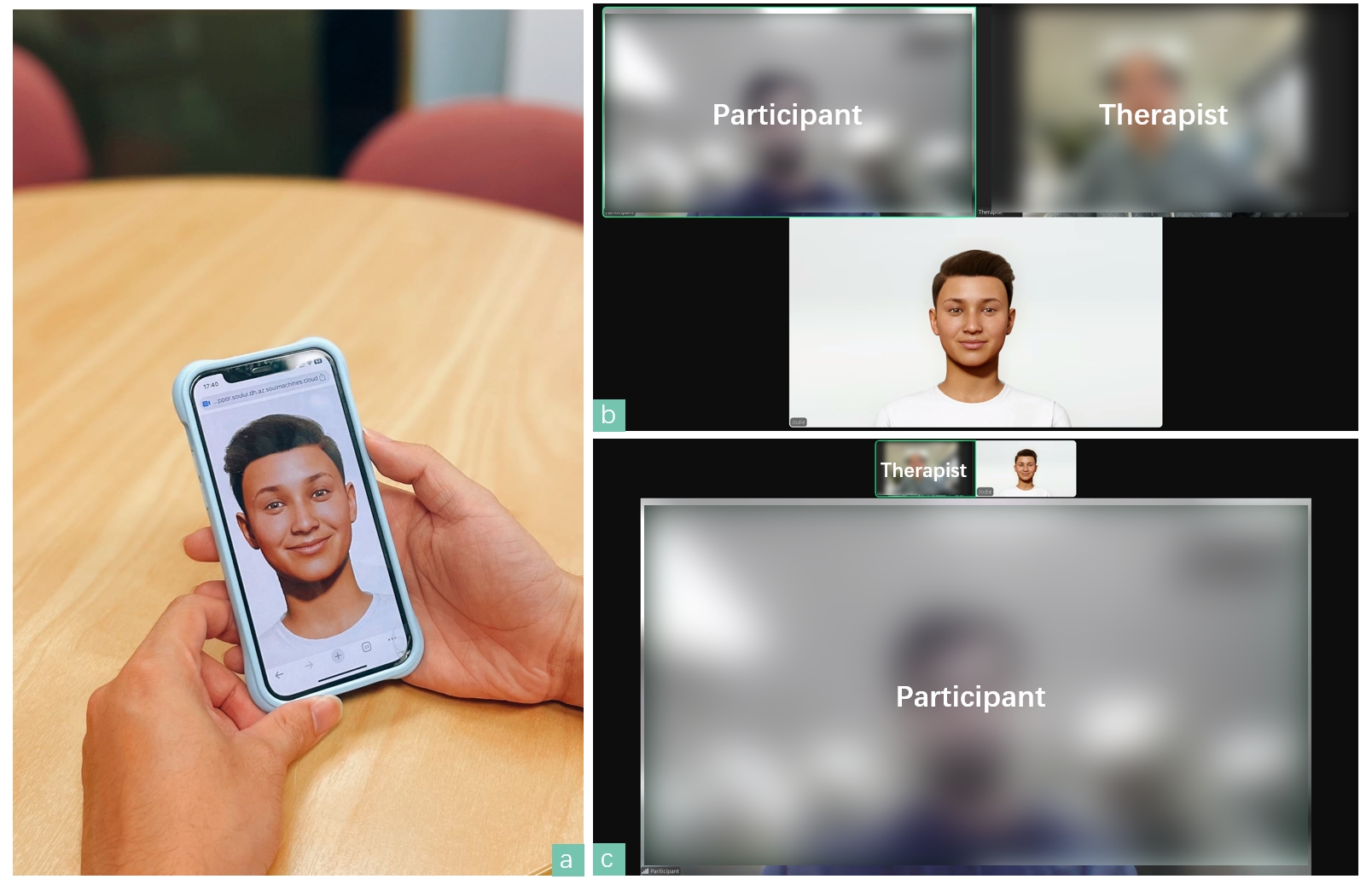}
    \caption{a: Daily mood journaling with Jodie. b: Jodie in Zoom session-Participants layout. c: Jodie in Zoom session-Therapist layout}
    \Description[Jodie's interface in daily and therapy mode]{Daily mood journaling with Jodie on mobile phone. Remote therapy with Jodie join Zoom session as supporter}
    \label{fig:jodie_use}
\end{figure}

This research received ethical approval from the [Blind for Review] University's Institutional Review Board (IRB) prior to participant recruitment. The ethics committee carefully reviewed our experimental design, therapist qualifications, participant safeguards, data management protocols, and informed consent procedures to ensure compliance with ethical standards for human subjects research.

\subsection{Participant Recruitment}
Since our experiment included a therapist-led therapy session, we first conducted therapist recruitment. To maintain consistency across sessions with different participants, we decided to have only one therapist conduct all therapy sessions, with the same session structure. We established specific selection criteria for choosing the therapist. First, the selected therapist needed to have substantial experience with human supporters in therapeutic contexts, as this experience would enable them to better understand and evaluate Jodie's supportive functions in comparison to human supporters. Second, the therapist needed to be well-versed in remote psychotherapy practices, given that our study was conducted entirely via Zoom. Third, the therapist needed to have availability throughout the anticipated study period to ensure consistent participation across all therapy sessions. After comprehensive consideration, we selected T1, who had participated in our Formative Study, to serve as the therapist for the user study. T1 is a clinical therapist with 3 years of experience working primarily with adolescents and adults. Her specializations include CBT and SFBT. She received specialized training in remote psychotherapy during her professional development and currently maintains both in-person and online client practices, making her well-versed in both face-to-face and remote therapeutic modalities.

After explaining our experimental objectives to the therapist, we jointly developed the following recruitment criteria:
\begin{enumerate}
  \item Participants have recently experienced feelings of depression or anxiety.
  \item Participants must be willing to participate in a remote psychotherapy session with a real therapist.
  \item Participants must not have been diagnosed with any clinical disorder.
  \item Participants must not be currently undergoing any type of psychotherapy.
  \item Participants must be fluent in both speaking and understanding English.
\end{enumerate}
These recruitment criteria were designed to maximize voluntary participation in remote psychotherapy while minimizing potential harm to participants during the experimental process.

Over a two-month recruitment period, we enrolled 16 participants, of whom 14 completed the experiment (two withdrew due to scheduling conflicts). The final participant group consisted of 7 males and 7 females, aged 19-28 years (mean=22.92, SD=2.63). The majority of participants (12, 85.7\%) are university students, and P8 and P14 are professional staff at the university. All participants scored within the minimal to moderate depression range on the PHQ-8 assessment \cite{kroenke2009phq}. PHQ-8 (Patient Health Questionnaire-8) is an 8-item widely-used and validated brief screening instrument for assessing depression severity. None of the participants had previous experience with psychotherapy, though three had attended job-related counselling sessions.

\subsection{Study Design and Procedure}
Our study was structured into four main phases: Study Briefing, Daily Mood Journaling, Remote Psychotherapy Session, and Feedback Collection.

\subsubsection{Study Briefing (15-20 minutes)}
At the beginning of the experiment, participants completed a demographic questionnaire. The researcher then used slides to introduce the overall experimental process and demonstrated how to interact with Jodie to complete mood journals. After addressing any questions, the researcher provided participants with support resources accessible during the study. These resources were made available in case participants experienced emotional distress or depressive symptoms, ensuring they knew where to seek help if needed. Finally, participants received the Jodie web application link via email.

\subsubsection{Daily Mood Journaling (7 days)}
Participants were required to use their personal devices to complete at least 7 days of mood journaling with Jodie (see Figure \ref{fig:jodie_use}a). While participants could exit their conversations with Jodie at any time, they needed to complete at least the mood scaling activity (selecting a number from 1-9) to be considered as having completed the day's mood journal. When participants reached the fifth day, the researcher contacted them via email to schedule their remote psychotherapy session with the therapist. Daily mood journaling continued until the day before the remote psychotherapy session.

\subsubsection{Remote Psychotherapy (single session, 20-25 minutes)}
To ensure the single remote therapy session is clinically appropriate and practically feasible within our experimental constraints, we invite T1 to help us design the session structure(see Appendix \ref{appendix-therapy}).T1 chose to focus on the Exploration Stage commonly used in early psychotherapy and optimized the process for our experimental setting. The entire session was conducted via Zoom. To minimize environmental influences on both the participant's and therapist's experience of the session, we required both to complete the session in designated experiment rooms.

Participants entered the Zoom meeting 5 minutes before the therapy session began. The researcher then connected a Zoom account featuring Jodie's virtual camera, allowing Jodie to conduct a pre-therapy check-in conversation with the participant (see \ref{therapyModeFeature}). The therapist subsequently joined the Zoom meeting to begin the formal therapy session, during which Jodie remained present. After the therapy session concluded, Jodie engaged the participant in a post-therapy reflection conversation before leaving the Zoom meeting.
\subsubsection{Feedback Collection (25-45 minutes)} 
After Jodie exited the meeting, the researcher entered the participant's experiment room and invited them to complete questionnaires and participate in a semi-structured interview.

After all participants completed the experiment, we invited the therapist to participate in an interview to gather her impressions and experiences throughout the entire process.

\subsection{Data Collection}
To comprehensively capture participants' experiences throughout the experiment and their perceptions of Jodie, we employed a mixed-methods approach combining quantitative measures and qualitative interviews.

For quantitative data, we selected the System Usability Scale (SUS) \cite{brooke2013sus} to evaluate Jodie's usability as a virtual agent, and the Session Evaluation Questionnaire (SEQ) \cite{stiles2002session} to assess participants' perceptions of the therapy session. Additionally, we utilized selected dimensions from the mobile Agnew Relationship Measure (mARM) \cite{berry2018assessing}, specifically focusing on the Openness, Bond, and Partnership subscales to measure the relationship between participants and Jodie. These dimensions were particularly relevant given Jodie's role as a virtual supporter in the therapeutic context.
For qualitative data, we conducted semi-structured interviews to gather participants' thoughts and experiences (see Appendix \ref{appendix-evaluation}). The interviews explored participants' overall impressions of Jodie, their experiences interacting with the agent during daily mood journaling, their perceptions of Jodie's presence during the therapy session, specific helpful or challenging aspects of the interactions, and suggestions for improvement. All interviews were conducted face-to-face and recorded with participant consent.

Since psychotherapy is a collaborative activity between therapist and client, we also collected the therapist's perspectives through a semi-structured interview. We believe that how the therapist perceives Jodie and her understanding of the therapy session process are equally important. After the last therapy session, we invited T1 to reflect on her experience, with particular focus on her perceptions of Jodie's presence and how these sessions compared to traditional remote therapy practice.

\subsection{Data Analysis}
Our data analysis employed a mixed-methods approach to integrate both quantitative and qualitative findings.

For quantitative data, we calculated descriptive statistics (means, standard deviations, ranges) for the SUS, SEQ, and mARM. These metrics provided a standardized assessment of Jodie's usability, participants' evaluation of the therapy session, and the quality of human-agent relationship that developed throughout the study.
For qualitative data, we collected 379 minutes (mean=27.07, SD=7.34) of interview data across all participants, and 36 minutes from an interview with T1. Then we applied thematic analysis to the transcribed interview recordings. First, two researchers independently familiarized themselves with the data and generated initial codes. These initial codes were then compared, discussed, and refined into a codebook. Using this codebook, all interviews were systematically coded, and the codes were clustered into meaningful themes and subthemes. Emerging themes were iteratively reviewed and refined to ensure they accurately represented the data while addressing our research questions. Throughout this process, we maintained an audit trail to ensure analytical rigor and conducted member checking with a subset of participants to validate our interpretations.

We then integrated the quantitative and qualitative findings to identify points of convergence and divergence, developing a comprehensive understanding of participants' experiences with Jodie in the therapeutic context and the factors influencing the effectiveness of AI-supported therapy approaches.

%% file: Sections/7-Results.tex
\subsection{Quantitative Results}

\subsubsection{System Usability}
Based on the SUS data, Jodie received a final score of 72.8 (SD=6.67), which corresponds to a Grade B- and an "Excellent" ranking according to the Curved Grading Scale for the SUS \cite{lewis2018system}. Analysis of individual question responses (see Figure \ref{fig:quat1}) revealed relatively high scores for Q3 "I thought the system was easy to use." (4.4/5), Q7 "I would imagine that most people would learn to use this system very quickly." (4.2/5) and Q9 "I felt very confident using the system." (4.5/5) These findings indicate that participants consistently found Jodie to be both easy to learn and easy to use. While the scores for Q1 "I think that I would like to use this system frequently." (3.7/5) and Q5 "I found the various functions in this system were well integrated." (3.6/5) were comparatively low, these results suggest that participants believe some improvements are needed for everyday use, which will be further explored through our interview findings.

\subsubsection{Session Perception}
\begin{figure}
    \centering
    \includegraphics[width=0.9\linewidth]{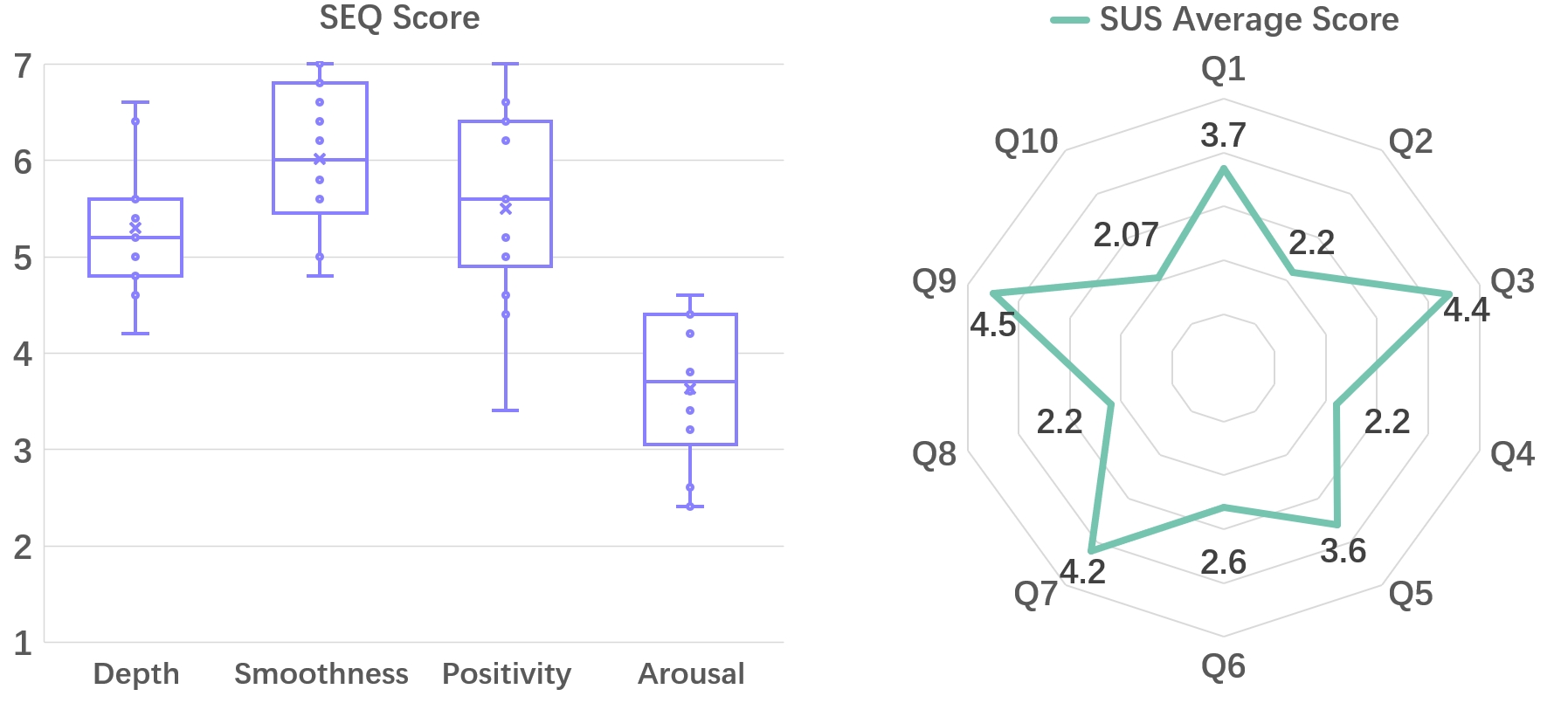}
    \caption{Session Evaluation Questionnaire scores and average scores for each System Usability Scale question}
    \Description[Result of System Usability Scale and Session Evaluation Questionnaire]{SUS got Grade B- and an "Excellent" ranking, SEQ result suggest participants experienced positive therapy session in general.}
    \label{fig:quat1}
\end{figure}

The SEQ results (see Figure \ref{fig:quat1}) provided a preliminary indication of how participants experienced the therapy session. Session depth averaged 5.3 out of 7 (SD=0.6) and smoothness averaged 6.01 out of 7 (SD=0.7). While these scores trend toward the positive end of the scale, they should be interpreted cautiously in the absence of a control condition or established benchmarks; nonetheless, the relatively low standard deviations suggest reasonable agreement among participants rather than polarized responses. Similarly, positivity averaged 5.5 (SD=0.9), and arousal averaged 3.63 (SD=0.7), consistent with a moderate level of emotional activation. Taken together, these results offer a tentative signal that participants did not find the session aversive, though the potential influence of demand characteristics cannot be ruled out. Qualitative findings from the interviews, reported in the following section, are used to triangulate and further contextualize these patterns.

\begin{figure}
    \centering
    \includegraphics[width=0.9\linewidth]{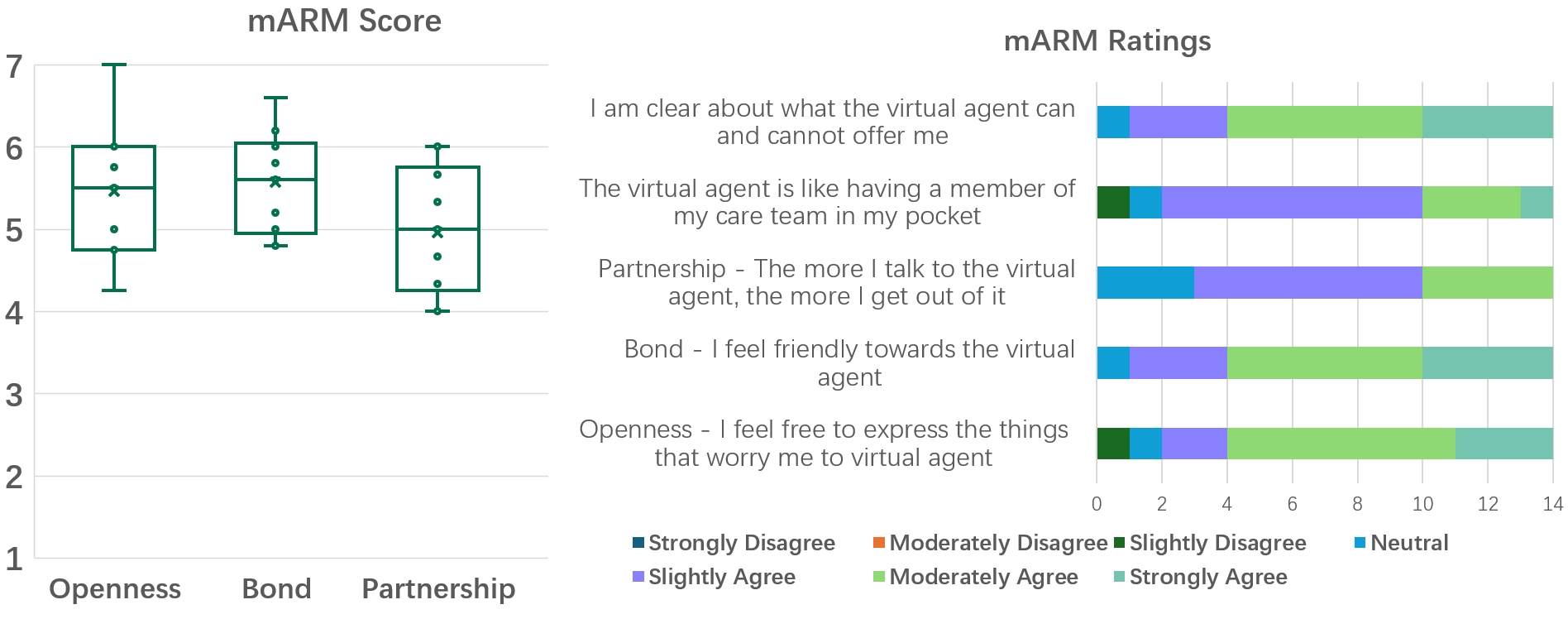}
    \caption{Openness, Bond and Partnership score, and selected questions' answers from mARM}
    \Description[Result of mARM]{Analysis of the mARM data revealed that participants rated their relationship with the agent positively across all measured dimensions}
    \label{fig:quat2}
\end{figure}

\subsubsection{Relationship towards Jodie}
Analysis of the mARM data (see Figure \ref{fig:quat2}) revealed that after one week of mood journaling and a therapy session with Jodie, participants rated their relationship with the agent positively across all measured dimensions: Openness (mean=5.46, SD=1.08), Bond (mean=5.57, SD=0.96), and Partnership (mean=4.95, SD=1.17) out of 7.

Two additional non-categorized questions provided further insights. For the statement "I am clear about what the virtual supporter can and cannot offer me," responses were predominantly on the agreeing end of the scale (1 neutral, 3 slightly agree, 6 moderately agree, 4 strongly agree), suggesting that most participants developed a reasonable understanding of Jodie's role boundaries over the study period. For the statement "The virtual agent is like having a member of my care team in my pocket," responses were more varied: 1 slightly disagree, 1 neutral, 8 slightly agree, 3 moderately agree, and 1 strongly agree, reflecting that while participants generally viewed Jodie as a useful resource, the extent to which it felt integrated into their care varied.

\subsection{Qualitative Results}
\subsubsection{Evolving from Task to Trust}
A notable pattern in participant experiences was the evolution of their relationship with Jodie from viewing interactions as obligatory tasks to engaging in genuine, trust-based sharing. Initially, most participants approached the daily mood journaling as a procedural requirement with limited personal disclosure. As P2 noted, \textit{"At first I just answered the questions without sharing much detail. It felt like completing a survey."} However, as interactions continued, participants described a shift toward more natural and motivated sharing. To triangulate participants' self-reported experiences, we examined behavioral data from the mood journaling sessions. Analysis of session duration across the week revealed a gradual increase in engagement over time. Session durations grew from an average of 3.07 minutes (SD=0.91) on Day 1 to 6.14 minutes (SD=1.95) by Day 7.

This transition was attributed to several factors, most prominently growing familiarity with Jodie and the perception of Jodie as a non-judgmental listener. P4 explained, \textit{"Later on I could share with her without any burden, because I felt I had become very familiar with her, and I knew she wouldn't judge me under any circumstances."} Several participants contrasted speaking with Jodie to written journaling, noting that verbalization offered a different quality of emotional processing: \textit{"Saying things out loud to Jodie feels different than writing them down. There's something about articulating thoughts that helps me understand them better."}, described by P1, who also noted to having a habit of diary writing.

The development of this trust relationship appeared to be cumulative, with each successful interaction building confidence for deeper disclosure in subsequent sessions. This finding supports our design goal of relationship building through daily interaction, demonstrating that consistent engagement can foster meaningful connections even with a virtual entity.

\subsubsection{Engagement During Different Therapy Phases}
Participants reported distinct experiences with Jodie during different phases of the therapy process, each offering unique benefits to the therapeutic experience.

Many participants indicated that seeing Jodie enter the Zoom room before therapy began created a sense of shared presence that reduced pre-session anxiety. \textit{"Seeing Jodie there made me feel like I wasn't alone in that waiting space."} (P7). Several noted that Jodie's presence triggered a readiness for sharing, a mindset carried over from their daily journaling experiences: \textit{"When I saw Jodie, I automatically shifted into a more open mindset, like 'okay, it's time to start sharing now.'"} (P12) This conditioning effect appeared to reduce anxiety about the up-coming therapy for most participants.

During the therapy session itself, the vast majority of participants reported that Jodie did not create any distraction and they could maintain focus on their conversation with the therapist. Some participants noted that Jodie's presence actually provided relief during moments of conversational anxiety: \textit{"There were times when I didn't know what to say next and felt awkward, but glancing at Jodie somehow made me feel less pressure."} (P9). Interestingly, we observed that all participants positioned Jodie's video frame between themselves and the therapist in the Zoom interface (see Figure \ref{fig:jodie_use}b), suggesting an intuitive placement of the supporter within the therapeutic relationship.

The post-therapy reflection conversations were characterized by most participants as a valuable transition phase. They described these interactions as helping them both process the therapy session and prepare to return to their personal lives. P5 noted, \textit{"Talking with Jodie afterwards helped me wrap up my thoughts from therapy and also gave me a moment to shift back into my regular mindset before jumping back into my day."}

\subsubsection{Desire for More Human-like Interaction}
While participants generally responded positively to Jodie, many expressed a desire for more sophisticated interaction capabilities and better dialogue control. During mood journaling, several participants wished for more empathic responses when sharing personal stories, rather than just attentive listening. As P2 suggested, \textit{"I'd like Jodie to engage more with what I'm saying, maybe ask follow-up questions or share observations rather than just acknowledging."} P6 described a particularly frustrating experience: \textit{"I was trying to tell her about something that happened at work that made me upset, but she kept saying she don't understand that. Although she went back normal after I restart the system, it really puts me off."} There are other participants reported similar experiences where Jodie's conversational boundaries disrupted emotional moments: \textit{"She'd be asking about my mood, and I'd try to explain like why I was feeling anxious, but she just keep asking me to pick a number. It felt like she was shutting me down."} (P9). P13 further noted that this rigid conversational pattern felt very unnatural and prevented the establishment of genuine connection, \textit{"After I became familiar with the process of chatting with Jodie, I felt like I wasn't talking to a person, but rather something like a talking questionnaire."}.

For therapy sessions, some participants hoped Jodie could actively participate in conversations when needed. P11 specifically mentioned wanting Jodie to help fill in information gaps: \textit{"I actually told Jodie about this on Wednesday, but when talking it again (to the therapist), I felt like I was forgetting something. It would be great if she could remember and help me there."} This suggests a potential role for Jodie as an information provider, one of the supporter functions identified in our formative study. Similar preferences emerged regarding post-therapy reflection, with participants expressing interest in Jodie offering her own observations rather than just soliciting their reflections. This desire for more active engagement suggests that as trust in the virtual supporter grows, users may seek more reciprocal interaction that more closely mirrors human supportive relationships.

\subsubsection{Friend vs. Assistant}
At the conclusion of interviews, we asked participants to describe their relationship with Jodie using three keywords, revealing an interesting bifurcation in how participants conceptualized the relationship. Two distinct perspectives emerged from this exercise.

The larger group (9 participants) used terms like "friend," "companion," and "caretaker" to characterize their relationship with Jodie. These participants described developing genuine social connections through Jodie's consistent listening presence. They expressed a desire to continue therapy sessions with Jodie, viewing the relationship as one of mutual growth: \textit{"It's about us growing together, I understand her better and she understands me better over time."} (P10). For these participants, Jodie had transcended her technological nature to become a social entity with whom they felt a meaningful connection.

The second group (5 participants) used more utilitarian descriptors such as "assistant," "tool," and "recorder." These participants maintained a more functional view of Jodie, describing her as "just a tool for providing emotional support." While they acknowledged the potential for deeper social connection with extended interaction, their current perception positioned Jodie primarily as a technological resource rather than a social companion. As one participant stated, \textit{"I appreciate what she does, but I don't think of her as a person or friend, she's more like a helpful app that's designed specifically for therapy support."} (P13).

This divergence in relationship models suggests that individual differences play a significant role in how users engage with virtual supporters. Some users may naturally anthropomorphize and develop social connections, while others maintain a more instrumental perspective. This finding has important implications for the design of virtual supporters, suggesting that flexibility in relationship models may be necessary to accommodate different user preferences and interaction styles.

%% file: Sections/8-Discussion_Implication.tex
\subsection{Benefits and Drawbacks of Jodie's Presence}
The results from our user study demonstrate that virtual supporters can provide meaningful therapeutic benefits, while also introducing unique challenges stemming from their technological constraints. Participants' experiences with Jodie reflected a complex interplay between meaningful supportive functions and significant limitations that affected both relationship development and therapeutic utility.

Participants reported that Jodie's presence in therapy sessions provided emotional grounding during moments of conversational anxiety. As P9 described, \textit{"There were times when I didn't know what to say next and felt awkward, but glancing at Jodie somehow made me feel less pressure."}. This suggests that virtual supporters may be able to create a similar idea of "a witnessing space" described by T8 in our design phase, which distributes the intensity of the one-on-one therapeutic encounter. Jodie's observer function provided valuable continuity through daily mood journaling, facilitated ongoing engagement that extended beyond therapy sessions. This benefit also aligns with the previous research on how social support could serve as a buffer against stressors and enhance psychological well-being \cite{feeney2015new}. 



However, these benefits were accompanied by significant drawbacks stemming from Jodie's rule-based dialogue system. The most frequently reported limitation was the inability to engage in natural, spontaneous conversation. When participants attempted to discuss topics outside the scripted mood journaling flow, they encountered generic redirect responses that disrupted rapport. Beyond conversational rigidity, participants noted that while Jodie demonstrated appropriate listening behaviors, her engagement lacked depth. P11 explained: \textit{"I told her about a really difficult conversation with my mom, and she just said 'I see' and nodded. I know she's programmed, but it felt hollow in that moment."} This gap between behavioral realism and empathic realism-responding appropriately to emotional content-became particularly apparent as participants developed expectations for reciprocal emotional engagement through daily interaction. These limitations represented deliberate design trade-offs for virtual agents in a therapeutic context \cite{fiske2019your}. The rule-based approach was intentionally chosen to maintain therapeutic boundaries, prevent over-reliance problems identified with human supporters, and ensure therapist-vetted content. However, our findings reveal that these protective constraints created new problems: the very features preventing Jodie from overstepping boundaries also limited her ability to fulfill supportive functions that participants naturally expected. 

\subsection{Mirrored Expectation: Virtual Supporter and Human Supporter}
Our findings reveal that users develop expectations for virtual supporters that strongly parallel those for human supporters, frequently using real-life relationship analogies to articulate their desires for Jodie's capabilities. These expectations evolved throughout the study, shifting from viewing interactions as obligatory tasks to more genuine, trust-based exchanges that mirrored familiar human relationships.

This is further supported by participants consistently referencing to existing human relationships when describing their desired improvements for Jodie. For example, P3 expressed wanting Jodie to provide advice during mood journaling because \textit{"that's how it is with my parents, I tell them about difficulties I'm facing, and then they offer me some suggestions."} This direct comparison to parental support demonstrates how participants naturally mapped familiar human supporter functions onto their expectations for the virtual agent. More broadly, participants' desired improvements for Jodie aligned remarkably with the human supporter behaviors identified in our formative study. They expressed a desire for Jodie to help supplement information during therapy sessions when they forgot details, similar to how human supporters often fill informational gaps. Some participants also desired Jodie to offer her perspective on issues being discussed, justifying this expectation with statements like \textit{"compared to the therapist, Jodie should understand me better."} reflecting the continuity of the relationship that human supporters typically provide.

These mirrored expectations highlight how users naturally transfer their mental models of human-to-human supportive relationships onto human-AI interactions, particularly in therapeutic contexts. Work from Holohan and Fiske \cite{holohan2021like} demonstrates how patients unconsciously project relationship frameworks from human interactions onto AI therapeutic agents. Similar research by Joseph et al. \cite{joseph2024transference} confirms that users experience transference with AI therapy tools, projecting feelings, expectations, and past experiences onto them in ways that mirror traditional therapeutic relationships. In our study, participants envisioned Jodie not just as a technological tool but as an intelligent entity capable of relationship-building, perspective-taking, and personalized support-qualities typically associated with human supporters.

\subsection{Continuous Presence as Relational Anchor: Bridging Daily Life and Therapeutic Encounters}
A distinctive feature of our virtual supporter design is Jodie's continuous presence across multiple therapy phases, from daily mood journaling through pre-therapy, in-session, and post-therapy interactions. We believe this extended accompaniment of Jodie across daily life and therapy sessions created a unique psychological transition effect that fundamentally altered how participants experienced the therapy encounter.

When participants saw Jodie appear in the Zoom room, the familiarity built through daily interactions manifested in what participants described as an automatic shift in mindset: \textit{"At first I was quite nervous, but when I saw Jodie come in, I felt like it wasn't a big deal, it was just chatting."} (P2). Rather than entering an unfamiliar, clinical interaction with a stranger, participants first encountered Jodie, which activated associations with the casual, non-judgmental interactions they had experienced throughout the week. This phenomenon can be explained by the CASA paradigm \cite{gambino2020building}, which suggests that people may mindlessly apply social heuristics for human interaction to computers. Jodie's presence served as a psychological bridge, transforming the potentially intimidating prospect of formal therapy into continuing an ongoing conversation with a trusted companion. This shift was also observed by the therapist, who described participants as having "lower self-defense mechanisms", which directly relates to high self-disclosure and trust levels in therapy. This finding aligns with previous research by Xiayu et al. \cite{chen2015drives}, which suggested that people can transfer their trust from a familiar target to another through certain interactions. In our study, introducing a familiar target, Jodie, into the initial therapy session appears to facilitate the establishment of trust between client and therapist more quickly and easily.


Therapeutic outcomes are determined not exclusively by in-session processes but are significantly influenced by between-session activities \cite{kazantzis2005using, kazantzis2010meta}. While our current study did not explore Jodie's role in between-session therapeutic work, based on feedback from post-therapy interactions, we can see the potential for Jodie to engage with clients in intersession activities. For example, P3 spontaneously inquired about continued access to Jodie following the experimental session, \textit{"Will Jodie still be available after this study? Can I talk to her after this?"}. She later revealed that she wanted to record her thoughts about the therapy session. This therapeutic reflection activity is considered particularly valuable for therapeutic consolidation  \cite{owen2012working}. Drawing upon the framework proposed by Bruna et al. \cite{oewel2024approaches} regarding technology-mediated support for intersession therapeutic activities, we believe that virtual supporters like Jodie could serve as consistent companions for therapeutic reflection and emotional processing between formal sessions.

\subsection{Therapist Perspective: Integrating Virtual Supporters into Clinical Practice} \label{Result_Therapist}
Psychotherapy is fundamentally a collaborative process between therapist and client \cite{horvath1993role}. Our therapist (T1) conducted 14 therapy sessions with Jodie present, providing valuable insights into how virtual supporters integrate into clinical practice from the professional perspective. Based on T1's experience, Jodie's presence did not cause meaningful disruption during therapy sessions. She attributed this to two key factors: the familiar Zoom interface and Jodie's appropriate behavioral boundaries. T1 explained that her standard practice of using Zoom's Speaker View and pinning the client's video meant that \textit{"Jodie was present but not prominent"}, occupying only a small portion of the screen (see Figure \ref{fig:jodie_use}c), similar to how a human supporter would sit in the background during in-person sessions. She appreciated that Jodie maintained appropriate boundaries by not speaking during therapy sessions, which \textit{"preserved the clarity of therapeutic roles"} and avoided the excessive support problem that therapists identified with human supporters in our formative study. \textit{"The fact that Jodie didn't interject or require any management from me was crucial. I could conduct the session as usual without having to navigate another voice in the conversation."} (T1).

Beyond noting the absence of disruption, T1 observed what she perceived as meaningful differences in how participants engaged during these sessions compared to her previous experiences with first-time clients. She reported that \textit{"participants seemed to have lower self-defense mechanisms based on my previous experience with first sessions."} providing professional validation for participants' self-reported feelings of comfort and reduced anxiety. This clinical observation of reduced defensiveness, suggests that Jodie's presence may have accelerated the trust-building process that typically requires multiple sessions. T1 also noted that participants demonstrated notable emotional fluency from the beginning: \textit{"There was a fluency in how they described their emotional states that you don't always see in first sessions. It's as if they had developed a vocabulary for their feelings through their interactions with Jodie."} This observation supports our design rationale for Daily Mode, suggesting that the week of mood journaling served dual purposes: establishing trust with Jodie while also preparing participants to articulate their experiences more effectively when meeting the therapist. The combination of reduced defensiveness and enhanced emotional articulation has important implications for therapeutic alliance formation \cite{ardito2011therapeutic}, as participants appeared to enter therapy with foundational skills and comfort levels that might otherwise take several sessions to develop.

\subsection{Boundaries and Restrictions for Virtual Supporters}
While participants desire virtual supporters to match or even exceed human supporters' capabilities, our interview with therapist during evaluation phase and relevant research highlight several significant risks in this approach.

First, the same challenges observed with human supporters-such as over-reliance and excessive support-can equally manifest with virtual supporters. This over-reliance has been identified as a common challenge across different AI mental health care tools \cite{martinez2022minding, fiske2019your}, where users may develop unrealistic expectations about virtual agents' therapeutic capabilities, potentially diminishing their motivation to seek appropriate human support or professional treatment. Therapists acknowledged that while enhanced functionality could provide additional benefits, they advocated for intentional limitations to establish clearer role boundaries. As T1 noted after the experiment, \textit{"More features don't always mean better therapeutic outcomes."} Therefore, beyond establishing clear role boundaries, preserving the virtual supporter's capacity to decline inappropriate requests becomes crucial. This capability mirrors an important aspect of healthy human relationships: the ability to set and maintain boundaries.

When examining the underlying mechanisms through which over-reliance between clients 
and supporters might develop \cite{overholser1990emotional, laestadius2024too}, it revealed the second ethical concern: emotional reliance between humans and AI agents. While our findings showed that participants who reported stronger emotional connections with Jodie demonstrated greater tolerance for system limitations and more positive experiences, it is important to differentiate between functional connections that support therapeutic goals (enabling self-disclosure and reducing anxiety) and emotional dependencies that may undermine wellbeing. As virtual supporters engage with users during periods of emotional vulnerability, individuals may be particularly susceptible to forming intense attachments and projecting feelings onto virtual agents in ways that mirror traditional therapeutic relationships \cite{fiske2019your}. In our study, P3's inquiry about continued access when the study ends hints at potential dependency formation. Martinez-Martin \cite{martinez2022minding} identifies how such dependencies can interfere with recovery by causing users to avoid addressing underlying issues requiring human intervention or experience deterioration in real-world social relationships. Thus, virtual 
supporters should be conceptualized and designed as "supportive technologies": supports therapeutic goals and complements human care, rather than "substitutive technology": displaces human connection and professional therapeutic relationships.

Another ethical concern when having a virtual supporter in mental health contexts is the risk of manipulation, especially when users try to seek advice from them. According to several research studies \cite{fiske2019your, haring2021robot, lucas2014s}, humans have been shown to be more compliant when a digital entity (e.g. a robot) asks them to do something or share private information as compared to with a person. This increased compliance is particularly concerning in psychotherapy-related scenarios, where users' vulnerability could mean a higher chance of exposing private information to virtual supporters. The power dynamics between AI systems and vulnerable individuals seeking mental health support can create an ethical dilemma that requires careful consideration. As observed in the ethics of care framework \cite{nicholson2019relational}, developers must acknowledge this potential for manipulation and implement safeguards that protect users while maintaining beneficial therapeutic interactions. This raises important questions about how we can design virtual supporters that provide genuine support without exploiting users' increased tendency to comply with machine-generated suggestions or recommendations.

These insights from professional and previous research emphasize the need for careful design of virtual supporter systems that complement rather than attempt to replace human connections, with clear frameworks for appropriate use and integration into therapeutic settings.

%% file: Sections/9-Limitation_FutureWork.tex
Our study has several limitations, with the first being a relatively small sample size (n=14) consisting of participants who had no previous formal psychotherapy experience. This homogeneity limits the generalizability of our findings to broader populations, particularly those with established therapeutic relationships or clinical mental health conditions. A similar limitation is present in our formative study; the participant group was predominantly female, which can limit the diversity of perspectives.

Similarly, the therapist results reported in Section \ref{Result_Therapist} are derived from only one therapist, although these findings are based on her experiences across 14 therapy sessions with Jodie present. While this provides depth of experience with the virtual supporter system, the single-therapist perspective limits the generalizability of our findings regarding how different therapists might perceive and integrate virtual supporters into their practice.

Another limitation is the short study period, which consisted of only 7 days of mood journaling and a single therapy session. While we identified several benefits in our results, this timeframe may not have been sufficient to observe long-term relationship dynamics or potential challenges that might emerge with extended use, such as how the virtual supporter should behave between therapy sessions.

Future research should include larger, more diverse participant groups, including individuals with varying levels of therapy experience and those with diagnosed mental health conditions who might have different needs and expectations from virtual supporters. And multiple therapists with varying backgrounds, therapeutic orientations, and levels of experience with remote therapy, to better understand the range of professional perspectives on virtual supporter integration.
Longitudinal studies spanning multiple therapy sessions would provide more comprehensive insights into how the client-virtual supporter relationship evolves over time and how it might influence therapeutic outcomes.

Additionally, controlled comparison studies are critical for establishing the specific value added by virtual supporters. Future work should employ designs that compare therapy sessions with virtual supporters present against control conditions (no supporter) and alternative configurations (e.g., human supporter), enabling researchers to quantify the impact of virtual supporter presence on therapeutic outcomes, client anxiety, and therapeutic alliance.
We also plan to explore LLM-driven virtual supporters, which could offer higher naturalness and better flexibility of interactions compared to rule-based dialogue management used in this study. However, we recognize that using LLMs means less controllability, which raises important questions about appropriate boundaries, potential therapeutic miscommunications, and ethical considerations in mental health support. These challenges will require careful design and implementation strategies to address effectively.

%% file: Sections/10-Conclusion.tex
This paper presents one of the first explorations of virtual supporters in remote psychotherapy, investigating how AI-driven agents can provide meaningful support during therapist-led sessions. Through our two-phase approach: formative study with therapists, design and development of Jodie, and user evaluation, we found that virtual supporters can provide meaningful psychological and functional benefits in therapeutic contexts. Jodie effectively fulfilled the "comforter" role by creating psychological safety and reducing anxiety, while serving as an "observer" through daily mood journaling that enhanced participants' emotional articulation. Notably, these functions manifested without disrupting the therapeutic process between client and therapist, addressing concerns about appropriate role boundaries in therapeutic contexts.

Our research contributes to the growing field of AI in mental healthcare by demonstrating that virtual supporters can serve as meaningful extensions of therapeutic environments, particularly in remote settings where traditional support networks may be less accessible. At the same time, our findings regarding users' diverse relationship models with Jodie-conceptualized either as "friend" or "assistant"-highlights the need for flexible, ethically-bounded approaches to virtual support. As mental healthcare continues to evolve toward hybrid delivery models, thoughtfully designed virtual supporters represent a promising direction for making therapy more accessible and comfortable, while respecting the essential human dimensions of the therapeutic process. 